\pdfoutput=1
\documentclass{aa}
\usepackage[pdftex]{epsfig,graphicx,color}

\usepackage{txfonts}
\begin{document}
   \title{Dynamic star formation in the massive DR21 filament}
   \author{N. Schneider \inst{1}
   \and  T. Csengeri     \inst{1}
   \and  S. Bontemps     \inst{2}
   \and  F. Motte        \inst{1}
   \and  R. Simon        \inst{3}
   \and  P. Hennebelle   \inst{4}
   \and  C. Federrath    \inst{5}
   \and  R. Klessen      \inst{5,6} 
           }
   \institute{Laboratoire AIM, CEA/DSM - INSU/CNRS - Universit\'e Paris 
        Diderot, IRFU/SAp CEA-Saclay, 91191 Gif-sur-Yvette, France
     \and 
        OASU/LAB-UMR5804, CNRS, Universit\'e Bordeaux 1, 33270 Floirac,
        France    
     \and  
        I.\,Physikalisches Institut, Universit\"at zu K\"oln,
        Z\"ulpicher Stra{\ss}e 77, 50937 K\"oln, Germany
     \and
        Laboratoire de radioastronomie, UMR CNRS 8112, \'Ecole normale sup\'erieure et 
        Observatoire de Paris, 75231 Paris, France  
     \and 
        Zentrum f\"ur Astronomie der Universit\"at Heidelberg, 
        Inst. f\"ur Theor. Astrophysik, Albert-Ueberle Str. 2, 69120 Heidelberg, Germany
      \and
        Kavli Institute for Particle Astrophysics and Cosmology, Stanford  
        University, Menlo Park, CA 94025, U.S.A.
     }

\offprints{N. Schneider}

\mail{nschneid@cea.fr}

\titlerunning{DR21 filament}

\authorrunning{N.~Schneider et al.}

\date{\today}


\abstract
{The formation of massive stars is a highly complex process in which
it is unclear whether the star-forming gas is in global
gravitational collapse or an equilibrium state supported by
turbulence and/or magnetic fields.}
{By studying one of the most massive and dense star-forming regions in
the Galaxy at a distance of less than 3 kpc, i.e. the filament
containing the well-known sources DR21 and DR21(OH), we attempt  
to obtain observational evidence to help us to discriminate between 
these two views.}
{We use molecular line data from our $^{13}$CO 1$\to$0, CS 2$\to$1,
and N$_2$H$^+$ 1$\to$0 survey of the Cygnus X region obtained with the
FCRAO and CO, CS, HCO$^+$, N$_2$H$^+$, and H$_2$CO data obtained with the IRAM
30m telescope.} 
{We observe a complex velocity field and velocity dispersion in the
DR21 filament in which regions of the highest column-density, i.e., dense
cores, have a lower velocity dispersion than the surrounding gas and
velocity gradients that are not (only) due to rotation.  Infall
signatures in optically thick line profiles of HCO$^+$ and $^{12}$CO
are observed along and across the whole DR21 filament. By modelling the
observed spectra, we obtain a typical infall speed of $\sim$0.6 km
s$^{-1}$ and mass accretion rates of the order of a few 10$^{-3}$
M$_\odot$ yr$^{-1}$ for the two main clumps constituting the filament.
These massive clumps (4900 and 3300 M$_\odot$ at densities of around
10$^5$ cm$^{-3}$ within 1 pc diameter) are both gravitationally
contracting. The more massive of the clumps,
DR21(OH), is connected to a sub-filament, apparently 'falling' onto
the clump. This filament runs parallel to the magnetic field.}
{All observed kinematic features in the DR21 filament (velocity field,
velocity dispersion, and infall), its filamentary morphology, and the
existence of (a) sub-filament(s) can be explained if the DR21
filament was formed by the convergence of flows on large scales and is
now in a state of global gravitational collapse. Whether this
convergence of flows originated from self-gravity on larger scales or
from other processes cannot be determined by the present study. The
observed velocity field and velocity dispersion are consistent with
results from (magneto)-hydrodynamic simulations where the cores lie 
at the stagnation points of convergent turbulent flows. } 


   \maketitle
%

\section{Introduction} \label{intro}

Our goal in this series of papers is to investigate the generic link
between the large-scale (several 10 pc) molecular-cloud spatial
structure (Schneider et al. 2010), medium-scale ($<$10 pc) dynamic
fragment properties (this paper), and the occurrence of high-mass star
formation on small (0.1-0.5 pc) scales (Bontemps et
al. \cite{bontemps2010}, Csengeri et al. \cite{csengeri2010}). To
achieve this, we make use of our multiwavelength study in Cygnus X
(see below). This region has already been shown to represent an
excellent laboratory for studies of high-mass star formation.

Molecular clouds (MCs) are the birthplaces of low- and high-mass
stars. Understanding their formation and evolution is essential to 
understanding in general the star formation (SF) process.  There are 
currently two different scenarios for the evolution of MCs and stars,
a {\sl dynamic} and a {\sl quasi-static} view.

In the dynamic context, MCs are transient, rapidly evolving entities
that are not in equilibrium, and the spatial and velocity structure of
the cloud is determined by compressible supersonic turbulence (see
e.g., Mac Low \& Klessen \cite{maclow2004} for an overview).  The
driving sources of turbulence can be diverse and their relative
importance remains a  subject of debate.  Large-scale mechanisms such as
supernovae explosions, should be the most importantXS, based on theoretical
arguments extensively discussed by MacLow \& Klessen
(\cite{maclow2004}).

Turbulence may also occur during the formation process of
molecular clouds (Vazquez-Semandeni et al. \cite{vaz2002}, Klessen \&
Hennebelle \cite{klessen2010}) within large-scale colliding flows of
mostly atomic gas in the galactic disk, generated by dynamic
compression in the interstellar medium or other large-scale
instabilities (see, e.g., Hennebelle \& Audit \cite{hennebelle2007},
Vazquez-Semadeni et al. \cite{vaz2008}, Hennebelle et
al. \cite{hennebelle2008a}, Banerjee et al. \cite{banerjee2009}). High
velocity compressive flows form dense structures at stagnation points
that may collapse to form stars/clusters (Ballesteros-Paredes et
al. \cite{ball2003}, Vazquez-Semadeni et al. \cite{vaz2007},
\cite{vaz2009}, Heitsch et al. \cite{heitsch2008}).  In this case, the
velocity field and velocity dispersion of molecular clumps may 
contain the signature of the external convergent motion {\sl and} the
compressive, gravitational contraction (this was already proposed by
Goldreich \& Kwan \cite{goldreich1974}). The star-forming core is then
the dense post-shock region with a more quiescent velocity dispersion
than the turbulent flow (Klessen et
al. \cite{klessen2005}). This scenario is valid for both, low- and
high-mass star formation.  However, there is no direct observational
confirmation of the existence of {\sl convergent flows}. Studying the
dynamics of HI and molecular gas may be a possibility (Brunt
\cite{brunt2003}), though it is not clear what observational
signatures are to be expected. An indirect argument for molecular
cloud formation out of colliding HI flows was provided by Audit \&
Hennebelle (\cite{audit2009}), who showed that their models closely
reproduce the observed clump mass spectra and Larson-relations. It has
also been noted that the lifetime of the cloud in the gravoturbulent
framework is short (one dynamical crossing time, i.e. $\sim$10$^7$ yr
for giant molecular clouds) and the star formation process is rapid.

In the quasi-static view (see McKee \& Ostriker
\cite{mckeeostriker2007} and references therein), MCs are formed by
large-scale self-gravitating instabilities such as spiral density
waves. (High-mass) star formation is approached in the 'turbulent
core' model (McKee \& Tan~\cite{mckee2003}) in which the star-forming
cores are supersonically turbulent. Turbulent magnetic and thermal
pressure supports the clump against self-gravity.  Because of energy
injection from newly formed stars, the clumps and most of the cores
may maintain their equilibrium, unless they quasistatically evolve to a
gravitationally unstable state to finally form stars (if the magnetic
flux diffuses out of the clump by ambipolar diffusion). In the case of
low-mass stars, the collapse of a rotating cloud of gas and dust leads
to the formation of an accretion disk through which matter is
channeled onto a central protostar.  For stars with masses higher than
about 8 M$_\odot$ this mechanism of star formation is less clear due
to the strong radiation field that pushes against infalling material
and may halt accretion. However, theoretical work (Yorke
\& Sonnhalter \cite{yorke2002}; Krumholz, McKee, Klein
\cite{krumholz2005}; Peters et al. \cite{peters2010}) has shown that
outflows lead to anisotropy in the stellar radiation field, which
reduce the radiation effects experienced by gas in the
infalling envelope. Thus, massive stars may therefore be able to form
by a mechanism similar to that by which low mass stars form. Peters et
al. (\cite{peters2010}) even show that clustered SF is a natural
outcome of massive SF even in the presence of radiative
feedback. However, clustered SF is also a natural result of the
{\sl competitive accretion} scenario (Bonnell \& Bate
\cite{bonnell2006}), in which the fragmentation of a turbulent cloud
produces stars with masses of the order of the Jeans mass within a
common gravitational potential.  These stars, located close to the centre
of the potential, accrete at much higher rates than isolated stars
and become massive.

Both scenarios have in common that massive ($>$10\,M$_\odot$) stars
form only from the coldest, and densest, cores (size scale $<$0.1 pc)
of molecular clouds and that high infall/accretion rates are required
to overcome feedback processes (ionizing radiation, jets,
outflows). In the dynamic framework, stars form as the dynamical
evolution of the MC progresses with gas continuously being accreted.
Individual clumps ($\sim$0.1--0.5 pc) fragment into hundreds of
protostars, possibly competing for mass in the central regions of the
cluster (Bonnell \& Bate \cite{bonnell2006}).  Support from magnetic
fields (Hennebelle \& Teyssier \cite{hennebelle2008b}) and/or
radiation (Krumholz \cite{krumholz2006}, Bate \cite{bate2009}) could
drastically limit the level of fragmentation and channel the global
infall to fewer, more massive protostars (Bontemps et
al. \cite{bontemps2010}).

From the observational point of view, indications of {\sl global
collapse} have been detected using molecular lines. Spectral profiles
of high density tracers, usually combining at least an optically thin
and an optically thick line, are good probes of infalling gas
(e.g. Myers et al. \cite{myers1996}).  Observations in low-mass
star-forming regions (e.g.  Lee et al. \cite{lee2003}) contain a
mixture of infall (blue-shifted emission and/or redshifted
self-absorption in the optically thick line) and outflow asymmetry.
This sort of line profile was also found in the high-mass star-forming
region W43 (Motte et al.~\cite{motte2003}). Peretto et
al. (\cite{peretto2006}) proposed that a peculiar velocity
discontinuity could be the result of some large-scale motion,
originating from self-gravity (Peretto et al. \cite{peretto2007}). The
theory of {\sl turbulent core formation}, on the other hand, predicts
the existence of massive pre-stellar cores that have not been detected
so far, not even in sensitive and extensive dust continuum studies
(e.g. Motte et al.  \cite{motte2007}).

In this paper, we shortly introduce the Cygnus X region and the DR21
filament in Sect.~\ref{cygnusx} and describe the molecular line
oberservations in Sect.~\ref{obs} and show maps and spectra in
Sect.~\ref{results}. Section~\ref{analysis} presents an analysis of
the kinematic structure and the physical properties of the DR21
filament and in Sect.~\ref{discuss} we use our findings to test the
conditions of the different high-mass star formation and turbulence
models. Section~\ref{summary} summarizes the paper.

\begin{figure*}[ht]
\includegraphics[angle=0,width=140mm]{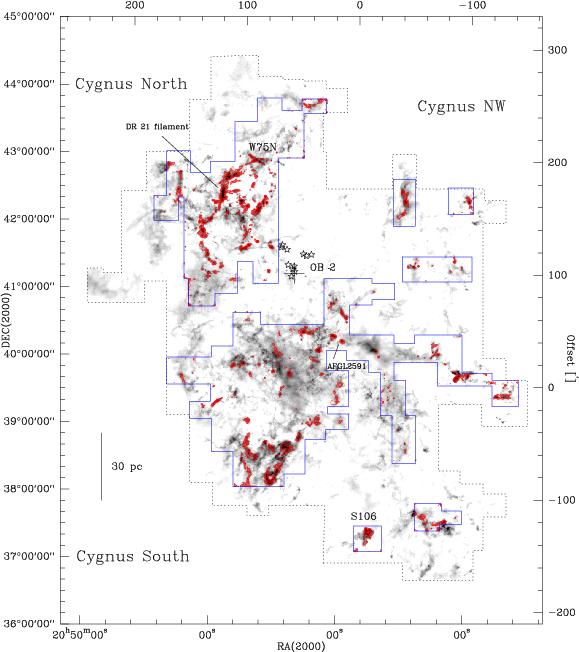}
\caption [] {Zeroth moment map of CS 2$\to$1 emission (red contours,
   mapping area indicated by a continuous blue line) in Cygnus X overlaid
   on a moment map of $^{13}$CO 1$\to$0 emission (mapping area
   indicated by a dashed line). Both maps were obtained with the
   FCRAO and have an angular resolution of $\sim$50$''$. Emission was
   integrated in the velocity range --10 to 20 km s$^{-1}$ and the
   grey scaling ranges from 1 to 15 K km s$^{-1}$. Red contour levels 
   for CS are at 0.5,1,2,4 K km s$^{-1}$. The DR21 filament is indicated, as
   well as some prominent objects in Cygnus X (S106, AFGL2591, W75N, 
   the most massive stars of the OB2 cluster). }
\label{overview}
\end{figure*}

\begin{figure*}[ht]
\includegraphics[angle=0,width=80mm]{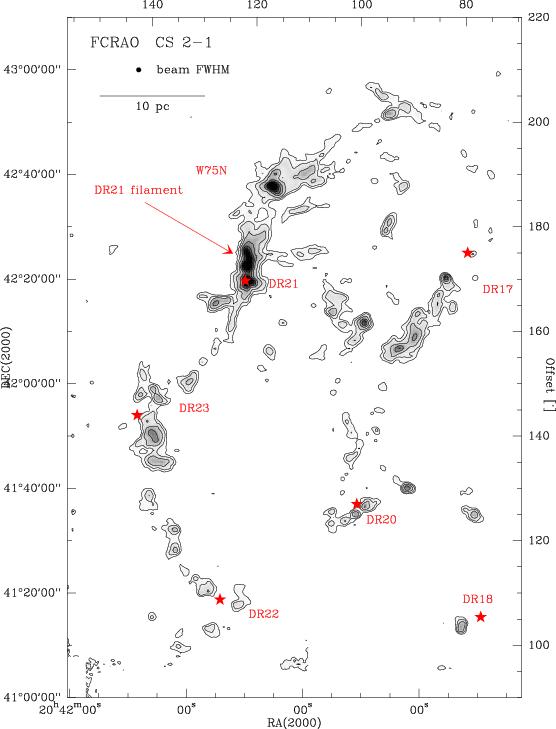}
\hskip 1cm \includegraphics[angle=0,width=80mm]{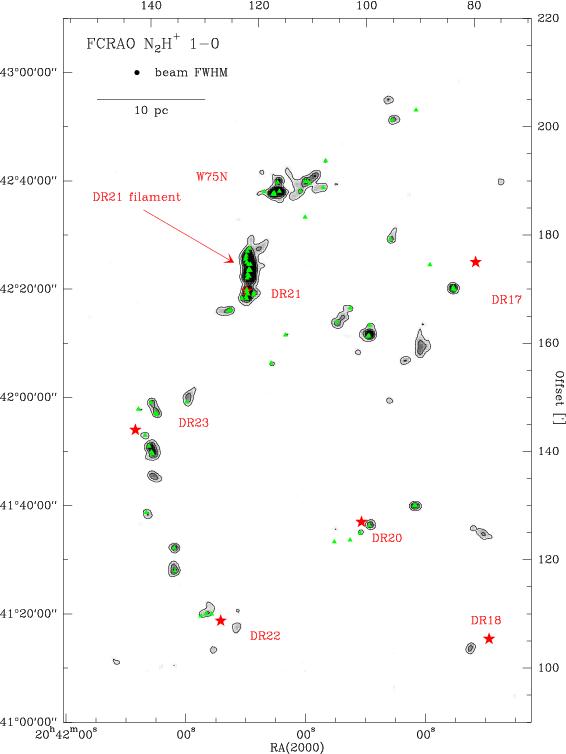}
\caption [] {Zeroth moment maps of CS 2$\to$1 (left, contour levels are 
0.5, 1, 2, 4 K km s$^{-1}$) and N$_2$H$^+$ 1$\to$0 (right, contour
levels are 0.5, 1.5, 3 K km s$^{-1}$) emission in Cygnus X North in
the velocity range --10 to 20 km s$^{-1}$ observed with the FCRAO.
Thermal HII regions (DR17--23) are indicated by red stars,
mm-continuum sources from Motte et al. (\cite{motte2007}) by green
triangles. The latter correspond very well with peaks of N$_2$H$^+$
emission.}
\label{fcrao-total}
\end{figure*}

\begin{table}[ht]  \label{obs-table} 
\caption{Observing parameters of the molecular line data obtained with the 
FCRAO and IRAM 30 m telescope: 
Column one and two indicate the molecular transition and frequency,
followed by the half power beam width (HPBW) in arcsec, $\eta_{\rm
mb}$ is the main beam efficiency, $T_{\rm sys}$ the system
temperature, $\Delta {\rm v}_{\rm res}$ denotes the velocity
resolution, and $\Delta T_{\rm rms}$ the average rms noise temperature
per channel on a $T_{\rm mb}$ scale.}
\begin{center}
\begin{tabular}{lcccccccc}
\hline \hline 
 & $\nu$ & \small{HPBW} & $\eta_{mb}$ & T$_{sys}$ & $\Delta {\rm v}_{\rm res}$ & $\Delta T_{\rm rms}$ \\
 & [GHz] & & & [K] & [km s$^{-1}$] & [K] \\
\hline 
{\bf FCRAO} & & & & & & \\
\hline
$^{13}$CO       1$\to$0 & 110.2  & 45$''$ & 0.48 & 210 & 0.067  & 0.48 \\
CS              2$\to$1 &  98.0  & 48$''$ & 0.48 & 200 & 0.075  & 0.38 \\
N$_2$H$^+$      1$\to$0 &  93.2  & 48$''$ & 0.48 & 200 & 0.079  & 0.38 \\
\hline                                               
{\bf IRAM} & & & & & & \\
\hline
$^{12}$CO       2$\to$1 & 230.8  & 11$''$ & 0.52 & 601 & 0.05  & 0.8 \\
H$_2$CO  \tiny{3(1,2)-2(1,1)} & 225.7  & 11$''$ & 0.55 & 549 & 0.05  & 0.65 \\
$^{13}$CO       2$\to$1 & 220.4  & 11$''$ & 0.57 & 580 & 0.03  & 0.9 \\
C$^{34}$S       2$\to$1 &  96.4  & 26$''$ & 0.77 & 210 & 0.025 & 0.5 \\
N$_2$H$^+$      1$\to$0 &  93.2  & 26$''$ & 0.77 & 158 & 0.03  & 0.1 \\
HCO$^+$         1$\to$0 &  89.2  & 28$''$ & 0.77 & 138 & 0.05  & 0.2 \\
H$^{13}$CO$^+$  1$\to$0 &  86.8  & 29$''$ & 0.78 & 114 & 0.025 & 0.25 \\
\hline
\end{tabular}
\end{center}
\end{table} 

\section {Cygnus X}  \label{cygnusx}
\noindent {\sl The Cygnus X region} \\
Cygnus X is one of the richest star-formation regions in the Galaxy,
covering an area of about 30 square degrees in the Galactic plane
around Galactic longitude 80$^\circ$ (Reipurth \& Schneider
\cite{reipurth2008}). It contains a prominent OB-association, Cyg OB2,
with $\sim$100 O-stars and a total stellar mass of up to 10$^5$
M$_\odot$ (Kn\"odlseder \cite{knoedl2000}).  From large-scale
$^{13}$CO 2$\to$1 (KOSMA\footnote{Cologne Observatory for
Submm-Astronomy}, Schneider et al. \cite{schneider2006}) and $^{13}$CO
1$\to$0 (FCRAO\footnote{Five College Radio Astronomy Observatory},
Schneider et al. \cite{schneider2007}, Simon et al., in prep.) 
surveys, we obtained a mass of a few 10$^6$ M$_\odot$ for the whole
molecular cloud complex that is divided into two parts -- Cygnus North
and South. These studies also showed that the majority of molecular
clouds is located at a common distance of about 1.7 kpc, i.e. the
distance of Cyg OB2.

Its proximity makes Cygnus X one of the rare laboratories where
different phases of massive star formation can be studied in detail.
More than 800 distinct HII regions, a number of Wolf-Rayet stars,
several OB associations, and at least one star of spectral type O4If
are known in Cygnus X, reflecting its record of high-mass star
formation in the past. Ongoing massive star formation is revealed by
wide-field (3 deg$^2$) 1.2 mm continuum imaging of the densest regions
in the Cygnus X molecular clouds (Motte et al. \cite{motte2007}). More
than 100 protostellar dense-cores were detected of which 40 are
likely to be the precursors of massive OB stars. High-angular
resolution observations with the Plateau de Bure Interferometer
(Bontemps et al. \cite{bontemps2010}) have indeed confirmed that the
most massive dense cores of this sample are the sites of massive star
formation, mainly in the form of clusters.  \\

\noindent {\sl The DR21 filament} \\
From our 1.2 mm continuum imaging and the $^{13}$CO surveys, we
concluded that the molecular ridge containing DR21 and DR21(OH)
(Dickel, Dickel \& Wilson \cite{dickeldickel1978}, Wilson \&
Mauersberger \cite{wilson1990}, Jakob et al. \cite{jakob2007}) is the
most active, dense (average density$\sim$10$^4$ cm$^{-3}$), and
massive (34 000 M$_\odot$) cloud in Cygnus X.

DR21 (Downes \& Rinehart \cite{downes1966}) itself is a group of
several compact HII regions (Harris et al. \cite{harris1973}) with an
associated, very energetic outflow (e.g. Garden et
al. \cite{garden1991a}, Russell et al. \cite{russell1992}), located in
the southern part of the ridge.  Three arcminutes further north lies
DR21(OH), famous for its maser emission (H$_2$O, Genzel \& Downes
\cite{genzel1977}; OH, Norris et al. \cite{norris1982}; CH$_3$OH,
Batrla \& Menten \cite{batrla1988}). Even further north lies the
massive star-forming region W75N (see Shepherd et
al. \cite{shepherd2004} for a review).  Very recently, the DR21
filament was again the target for a wealth of studies in various
wavelengths. Infrared images at 3.6, 4.5, and 8 $\mu$m from the
Spitzer satellite (Marston et al. \cite{marston2004}, Hora et al.
\cite{hora2009}) show the complexity of the region with many
IR-filaments perpendicular to the ridge that seem to be sites of star
formation (Kumar et al. \cite{kumar2007}). Dust continuum studies
(Vall\'ee \& Fiege \cite{vallee2006}, Motte et al. \cite{motte2007})
confirmed the detection of 8 compact, dense cores in the DR21 filament
(Chandler et al. \cite{chandler1993}) and found additional ones. Some
of them are high mass protostars and drive outflows detected in SiO
(Motte et al. \cite{motte2007}).  An H$_2$ 1-0 S(1) image
at 2.121 $\mu$m (Davis et al. \cite{davis2007}) shows at least 50
individual outflows driven by embedded low-mass stars.

\begin{figure*}[ht]
\includegraphics[angle=-90,width=150mm]{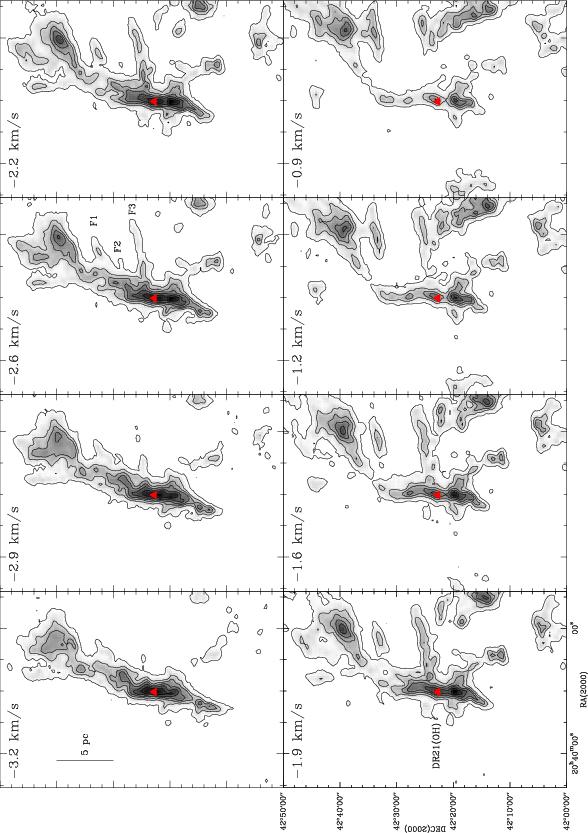}
\caption [] {Channel maps of $^{13}$CO 1$\to$0 emission in Cygnus X North in
  the velocity range --3.2 to --0.9 km s$^{-1}$ observed with the FCRAO.
  The red triangle shows the position of DR21(OH). The three 
  major subfilaments (F1, F2, F3) linked with the large DR21 filament are 
  indicated in the plot of velocity --2.6 km s$^{-1}$.}
\label{fcrao-channel}
\end{figure*}

We studied this particular filament in high-angular resolution
observations with the IRAM\footnote{Institut de Radio-Astronomie
Millimetrique} 30 m telescope in different molecular tracers
(isotopomeric CO lines, CS, C$^{34}$S, HCO$^+$, H$^{13}$CO$^+$,
N$_2$H$^+$, H$_2$CO) to investigate the distribution of the different
phases of molecular gas (cold dense cores, warm envelopes) and to
uncover infall/outflow signatures. The lower angular resolution
FCRAO data delineate the large-scale structure in which the DR21
filament is embedded.  A more detailed analysis of the FCRAO $^{13}$CO
data set is presented in Schneider et al. (\cite{schneider2007}, 
\cite{schneider2010}) and the CS and N$_2$H$^+$ data will be
discussed in a forthcoming paper. All molecular line data we obtained
serve as a basis for a physical model of the DR21 filament, including
detailed non-LTE line modelling. 

\section{Observations} \label{obs}
\subsection{FCRAO} \label{obs-fcrao}
We used data from a molecular line survey ($^{13}$CO and C$^{18}$O
J=1$\to$0, CS J=2$\to$1, and N$_2$H$^+$ J=1$\to$0) of the entire Cygnus
X region (Simon et al., in prep., Schneider et
al. \cite{schneider2007}) taken with the FCRAO 14m telescope.  The data were obtained
between 2003 December and 2006 January using the 32 pixel 'Second
Quabbin Optical Imaging Array' (SEQUOIA) in an On-the-Fly (OTF)
observing mode.  In this paper, we employ $^{13}$CO 1$\to$0, CS
2$\to$1, and N$_2$H$^+$ 1$\to$0 data that cover an area of 35 square
degrees in $^{13}$CO and 12 square degrees in CS/N$_2$H$^+$ on a
22$''$.5 grid. The beamwidth of the FCRAO at 93 (110) GHz is
45$''$(48$''$) and the dual channel correlator was configured to
obtain a velocity resolution of $\sim$0.1 km s$^{-1}$. The spectra
have a mean 1 $\sigma$ rms noise level of $\sim$0.2 K on a T$_A^*$
antenna temperature scale, i.e., not corrected for the main beam
efficiency of $\sim$0.48.  Pointing and calibration were checked
regularly at the start of the Cygnus observing interval and after
transit (no observations were performed at elevations higher than
75$^\circ$).  Pointing sources were SiO masers of evolved stars, i.e.,
$\chi$-Cyg, R-Leo and T-Cep, depending on LST-time. The calibration
was checked regularly on the position of peak emission in DR21 and
found to be consistent to within 10\%.

\subsection{IRAM} \label{obs-iram}
All molecular line maps\footnote{The IRAM data are available on request, contact 
nschneid@cea.fr.} obtained with the IRAM 30m telescope in the Sierra
Nevada/Spain were performed in June 2007.  Table~\ref{obs-table} gives
an overview of all lines and observational parameters. In order to
efficiently map an area of $\sim$4$'\times$10$'$, covering the whole
DR21 filament, we observed in the 'on-the-fly' mapping mode. The total
area was divided into 7 sub-maps that were scanned horizontally.  For
strong lines (e.g. CO, HCO$^+$), only one or two coverages were
required while weaker lines were observed more often or with two
receivers at once to obtain a sufficient signal-to-noise ratio. The
resulting regridded (5$''$) maps show slight smearing effects and
fluctuations in the intensity calibration only in the lines of
$^{12}$CO and H$_2$CO.  The center position of all maps is within
3$''$ of the MM2-position given in Mangum, Wooten, and Mundy
(\cite{mangum1991}): RA(2000)=20$^h$39$^m$00$^s$ and
DEC(2000)=42$^\circ$22$'$44$''$.  As OFF-position, we used a position
with offset 0$''$,1800$''$ from DR21(OH) where $^{12}$CO 2$\to$1
emission of around 1 K is observed at velocities around --42 km
s$^{-1}$, separated well from the bulk emission of the DR21
filament. All data were reduced using
GILDAS\footnote{http:iram.fr/IRAMFR/GILDAS}.
 
The spectral lines were recorded simultaneously in uniform, good
weather conditions (average atmospheric opacity of 0.2 at 230 GHz)
prevailing over several days.  We used the IRAM facility receivers and
autocorrelator, which were adapted in terms of spectral resolution and
bandwidth such that all lines have a similar velocity resolution
between 0.03 and 0.05 km s$^{-1}$.  Pointing and focus were checked
every 2 hours. The pointing accuracy was found to be better than 4$''$
and the receivers were aligned to within 2$''$.

\section{Results} \label{results}
\subsection{The DR21 filament as part of a complex network} \label{fcrao}
Figure~\ref{overview} shows the zeroth moment map of line integrated
(--10 to 20 km s$^{-1}$) CS 2$\to$1 emission obtained with the FCRAO
overlaid on a grey scale map of $^{13}$CO 1$\to$0 emission. The moment
maps were constructed following the masked moment procedure described
in Adler et al. (\cite{adler1992}). For CS and N$_2$H$^+$, we used a
temperature threshold of 0.1 K, for $^{13}$CO a threshold of 0.2
K. For the CS mapping, we focussed on the high column density regions
seen in $^{13}$CO. We adopt the nomenclature 'Cygnus-X North, South,
and Northwest', established in Schneider et al. (\cite{schneider2006})
for the different cloud regions.

The CS emission is more compact than that of $^{13}$CO and 
clearly delineates the high density regions. Interestingly, regions
of peak emission in $^{13}$CO do not always show up in CS. This
indicates that, though the column density is high, the average density
in these regions may stay below the critical density (n$_{cr}$) of the
CS line (a few 10$^5$ cm$^{-3}$). Abundance variations caused by chemical
effects probably also play a role. While in Cygnus North, and
partly in its north-west region, CS clumps form a connected network of
filamenatry structures, the emission distribution in Cygnus South is
more dispersed with more isolated CS clumps.

In the following, we concentrate on a discussion of the filaments
in Cygnus-North.  For that, we display in Fig.~\ref{fcrao-total} 
maps of CS 2$\to$1 and N$_2$H$^+$ 1$\to$0 emission where
the 'organization' of clumps along filaments is clearly visible. Even
the N$_2$H$^+$ map -- which traces only dense
(n$_{cr}$=0.5--70$\times$10$^6$ cm$^{-3}$) and {\sl cold} gas
(e.g. Tafalla et al. \cite{tafalla2002}) -- shows the same arrangement
of clumps. Most prominent in all maps is the filament containing the
HII region DR21 (see below).  The moment maps (Fig.~\ref{fcrao-total})
show that all evolved star formation sites in Cygnus North (indicated
by the HII regions DR17 to DR23) as well as future star-forming sites
(indicated by the mm-continuum sources tracing massive dense cores,
Motte et al. \cite{motte2007}) lie on filamentary structures.

\subsection{The DR21 filament} \label{iram}
From the large-scale, medium-angular-resolution ($\sim$1$'$) molecular
line maps (Sect.~\ref{fcrao}) and previous studies
(Sect.~\ref{intro}), the DR21 filament clearly appears to be the most
prominent star-forming region in Cygnus X. The DR21 region in the
south contains a very powerful, evolved outflow, while further north,
the DR21(OH) region contains maser sources, indicative of recent
star-formation activities. The mm-contiuum survey of Motte et
al. (\cite{motte2007}) revealed more than 20 protostellar objects,
i.e. sites of star formation. The filament is shown in more detail in
channel maps of $^{13}$CO 1$\to$0 emission (Fig.~\ref{fcrao-channel}),
obtained with the FCRAO. It reveals the characteristic N-S oriented
ridge known from other molecular line and continuum surveys (Dickel et
al. \cite{dickel1978}, Wilson \& Mauersberger \cite{wilson1990},
Vall\'ee \& Fiege \cite{vallee2006}, Motte et al. \cite{motte2007},
Jakob et al. \cite{jakob2007}). What is special in these channel maps
is the appearance of several E-W oriented subfilaments (indicated as
F1--F3 at v=--2.6 km s$^{-1}$) that are connected to the main DR21
filament and extend over several velocity channels. The most prominent
one (F3) has a length of around 10$'$ (5 pc) and is also visible in
the CS map (Fig.\ref{fcrao-total}) and in mm-continuum (Motte et
al. \cite{motte2007}), indicating that gas of high density is present.

\begin{figure}[ht]
\includegraphics[angle=-90,width=80mm]{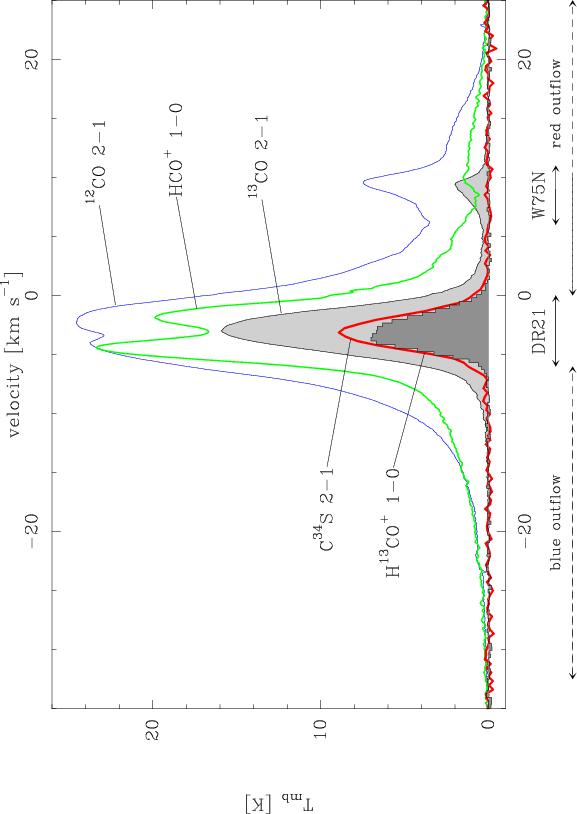}
\caption [] {Positionally averaged (across the whole observing region for 
each line) spectra of all lines observed in 
  the DR21 filament using the IRAM 30m telesope. The  HCO$^+$ line is multiplied by a factor
  8.5, the C$^{34}$S and H$^{13}$CO$^+$ lines by 10.} 
\label{average-spectra}
\end{figure}
 
\subsubsection{Average spectra} \label{spectra}
To study in more detail the physical properties (mass,
density, temperature) and kinematics (infall and outflow signatures,
velocity gradient) of the filament, we mapped the central part of the
filament containing DR21 and DR21(OH) in various molecular line
tracers at an angular resolution of between 12$''$ and 30$''$ using the
IRAM 30m telescope.

To distinguish the different emission features of the filament, we
produced positionally averaged spectra of five of the observed lines
(Fig.~\ref{average-spectra}). The averaging was performed across each  
map extent, i.e. covering the whole filament. The CO lines exhibit two velocity
components, one centered on --3 km s$^{-1}$ and one on +9 km
s$^{-1}$. They correspond to the molecular clouds associated with DR21
and W75N, respectively (see also Dickel et
al. \cite{dickel1978}). Though the W75N region itself is not included
in the mapping region, molecular gas related to this star-forming
region northwest of the DR21 filament (see Fig.~\ref{fcrao-total}) is
still present across the filament. However, the gas is more diffuse
and of lower density since a line at $+$9 km s$^{-1}$
(Fig.~\ref{average-spectra}) can only be seen in the low-density
tracers ($^{12}$CO and $^{13}$CO).  The HCO$^+$ line reveals a narrow
absorption feature at +9 km s$^{-1}$ (seen also by Nyman
\cite{nyman1983}), while the optically thin lines show no significant
emission for the W75N component.

The bulk emission of the DR21 filament is found at --3 km s$^{-1}$
(single Gaussian lines in C$^{34}$S, H$^{13}$CO$^+$, and
$^{13}$CO). The optically thick lines of $^{12}$CO and HCO$^+$ show
prominent, extended blue and red wing emission. The red wing, however,
is blended with the W75N component, seen in {\sl emission} in
$^{12}$CO and {\sl absorption} in HCO$^+$. The wings are caused by the
powerful outflow of DR21 (Garden et al. \cite{garden1991a}) but also
other outflow sources discussed in Sect.~\ref{outflows}. The
self-absorption feature at --3 km s$^{-1}$ is seen in the optically
thick lines of $^{12}$CO and HCO$^+$, while optically thin lines peak
at the velocity of the absorption dip. This may be a signature of
inflowing gas and is discussed in more detail in Sect.~\ref{infall}.

\begin{figure*}[]
\includegraphics[angle=-90,width=140mm]{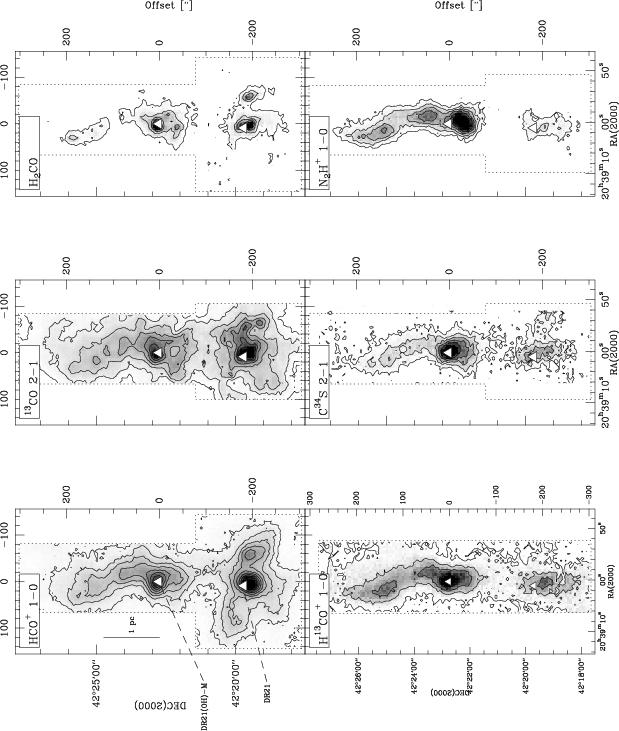}
\caption [] {Velocity-integrated spectral line maps of the DR21 filament. The HCO$^+$
  1$\to$0 and $^{13}$CO 2$\to$1 lines are integrated over a velocity range from --11 to
  5 km s$^{-1}$, the H$_2$CO 3(1,2)$\to$2(1,1) line from --11 to 2 km s$^{-1}$, and all other
  lines from --7 to 1 km s$^{-1}$. Contour levels follow the notation
  start/end/step with 15$\sigma$=3.8/87.3/7.5 K km s$^{-1}$ (HCO$^+$),
  6$\sigma$=1.3/10.9/3.9 K km s$^{-1}$ (H$^{13}$CO$^+$),
  30$\sigma$=34.1/249.4/23.9 K km s$^{-1}$ ($^{13}$CO), 6$\sigma$=5.9/14.7/2.9
  K km s$^{-1}$ (C$^{34}$S), 9$\sigma$=6.4/49.1/10.7 K km s$^{-1}$ (H$_2$CO),
  3$\sigma$=0.27/1.9/0.27 K km s$^{-1}$ (N$_2$H$^+$). Triangles indicate the
  positions of DR21(OH) and DR21.}
\label{maps1}
\end{figure*}

\begin{figure*}[ht]
\includegraphics[angle=-90,width=80mm]{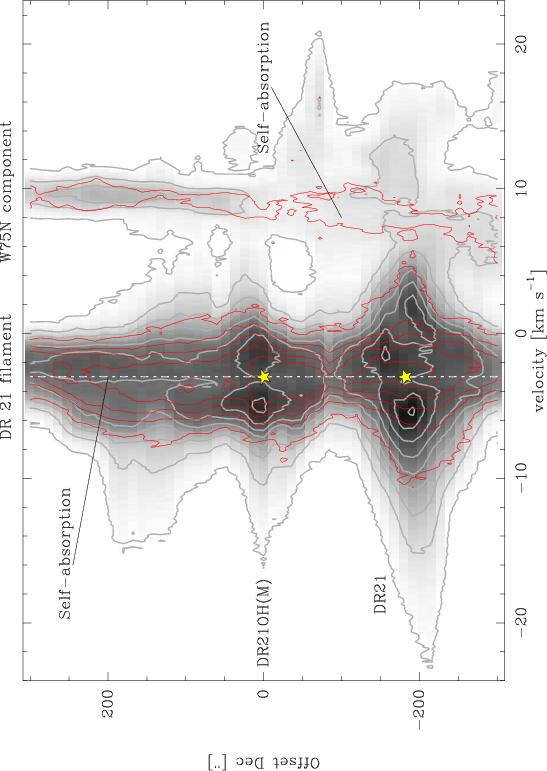}
\includegraphics[angle=-90,width=80mm]{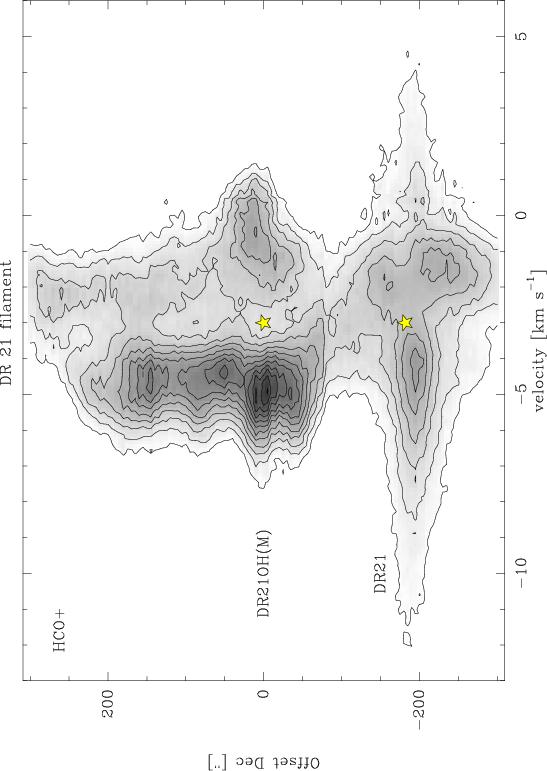}
\caption [] {Position-velocity cut along constant RA, i.e. along the
  DR21 filament. {\bf Left:} $^{12}$CO 2$\to$1 (grey scale and grey contours) and $^{13}$CO
  2$\to$1 (red contours). Contour levels are 2 to 34 K km s$^{-1}$ by
  2 K km s$^{-1}$ for $^{12}$CO and 1.5 to 16.5 K km s$^{-1}$ by
  3 K km s$^{-1}$ for $^{13}$CO. The stars mark DR21(OH) and DR21. 
  {\bf Right:} The same plot for HCO$^+$ with contour levels from 1 to 8 K km s$^{-1}$ by
  0.5 K km s$^{-1}$.}  
\label{pv}
\end{figure*}

\subsubsection{The DR21 filament in velocity-integrated maps} \label{maps}
Based on the average spectrum (Fig.~\ref{average-spectra}) we produced
line integrated maps of the bulk emission of the DR21 cloud
(Fig.~\ref{maps1}).  Though there are differences in the detailed
emission distribution, one can separate three main regions: a southern
cloud with DR21, a middle emission peak with DR21(OH), and a northern
extension with two secondary peaks.

While DR21 is most prominent in the HCO$^+$ and $^{13}$CO lines,
DR21(OH) is more pronounced in all optically thin lines. A small shift
in the position of peak emission of DR21(OH) can be discerned by
comparing C$^{34}$S/H$^{13}$CO$^+$ and N$_2$H$^+$. For N$_2$H$^+$, we
determined the optical depth using the hyperfine structure pattern of
this line and found that $\tau$ remains below 1 basically everywhere
in the map. The latter traces cold, dense gas that is located more
south of the warmer central part of the DR21(OH) region. Mauersberger
et al. (\cite{mauers1986}) report a temperature above 200 K in a small
region of size $<$9$''$ in DR21(OH), and Wilson \& Mauersberger
(\cite{wilson1990}) determine a rotation temperature of 34 K from
NH$_3$ observations. Mangum et
al. (\cite{mangum1991},\cite{mangum1992}) interpreted their
observations as a detection of a B-star within the DR21(OH) clump.
The northern part of the filament is more prominent in N$_2$H$^+$ and
H$^{13}$CO$^+$ emission and not in C$^{34}$S. Wilson \& Mauersberger
(\cite{wilson1990}), and Wienen (\cite{wienen2008}) derive values
around 20 K for this region.

Figure~\ref{pv} is a position-velocity cut in $^{12}$CO and
$^{13}$CO 2$\to$1 and HCO$^+$ 1$\to$0 along the filament in which all
data points in RA were averaged. The greyscale image of $^{12}$CO
emission shows that the bulk emission of the DR21 filament at --3 km
s$^{-1}$ and the bulk emission of the W75N component at 9 km s$^{-1}$
are apparently connected.  The gas linking the two components has a
low column density since it is only visible in $^{12}$CO and
not in $^{13}$CO (red contours). A lack of $^{12}$CO emission exists
between $\Delta
\delta$=0$''$ and --150$''$ at $\sim$8 km s$^{-1}$, indicating that
the (cooler) W75N cloud lies in front of the DR21 filament and causes
the $^{12}$CO self-absorption. Dickel et al. (\cite{dickel1978}) and
Bieging et al. (\cite{bieging1982}) came to the same conclusion
following their detection of self-absorbed line profiles in H$_2$CO
and OH toward DR21. However, the situation is slightly more
complicated because the W75N line component is blended in its southern
part ($<$0$''$) with emission from the 'Great Cygnus rift', which is
basically not seen in $^{13}$CO emission (Schneider et
al. \cite{schneider2007}). The rift is an extended, low-density gas
cloud at a distance of around 600 pc. Its emission can start as low as
+4 km s$^{-1}$ and makes it thus difficult to disentangle the
different emission features. It is, however, possible that there is a
real physical connection/interaction between the DR21 filament and the
W75N cloud, as proposed by Dickel et
al. (\cite{dickeldickel1978}). They suggested that the 'W75N' cloud is
differentially slowed down as it passes the edge of the DR21 cloud due
to the observed N-S oriented velocity gradient from +11 km s$^{-1}$ to
+9 km s$^{-1}$ (which is clearly seen in our CO data as well).

\begin{figure*}[ht]
\includegraphics[angle=0,width=70mm]{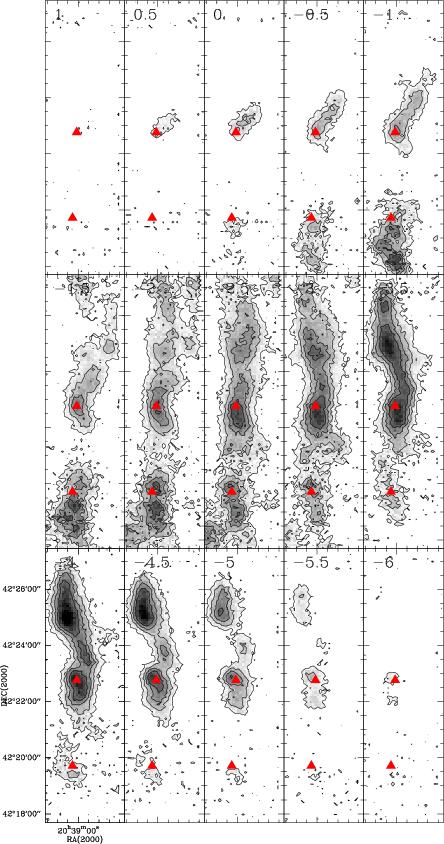}
\includegraphics[angle=0,width=70mm]{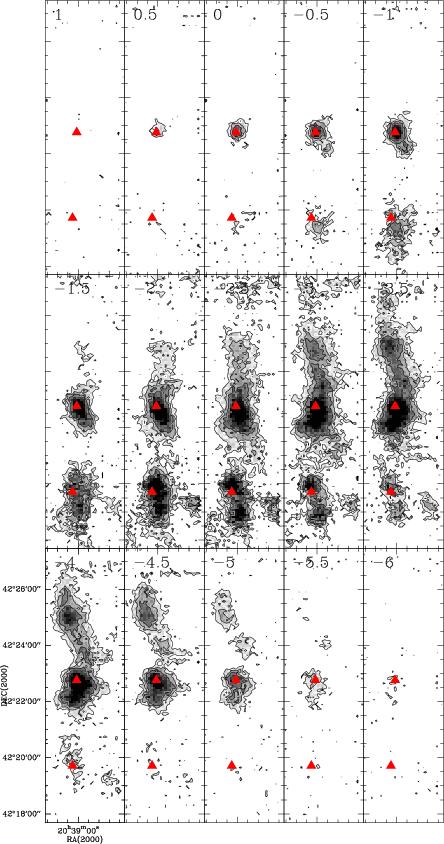}
\caption [] {Channel maps of H$^{13}$CO$^+$ 1$\to$0 (left) and C$^{34}$S 2$\to$1 
  emission (right) between +1 km s$^{-1}$ and --6 km s$^{-1}$. 
  The red triangles indicate the positions of DR21(OH)(north) and 
  DR21 (south), respectively.}  
\label{channel-h13co}
\end{figure*}

Several outflows are identified out of which the DR21 outflow at $\Delta
\delta$=--200$''$ is the most pronounced. However, its red flow is not
clearly defined because (i) it is blended with the 'Great Rift' emission
and (ii) because it runs into an HII region (Garden et
al. \cite{garden1992}). The blue flow, however, is clearly
defined. Other 'flow-features' are identified at $\Delta
\delta$=0$''$ (DR21OH(M)) and at $\sim$180$''$, both look well-defined at
negative velocities and disturbed for positive velocities. The
position-velocity diagram for HCO$^+$ shows that these outflows are
much less obvious than that towards DR21. We come back to outflows in
Sect.~\ref{outflows} and in particular the one seen at DR21OH(M) in
Sects.~\ref{noflow} and \ref{infall}.

Another feature visible in Fig.~\ref{pv} is a region of
self-absorption in $^{12}$CO ({\sl emission regions} in $^{13}$CO) at
around --3 km s$^{-1}$ along the whole filament (indicated by a dashed
white line). This central region of self-absorption is even more
clearly identified in the HCO$^+$ map. The strongest self-absorption
is found at the declination offsets of DR21 and DR21OH(M). We discuss
this finding in more detail in Sect.~\ref{infall}.

\subsubsection{A subfilament at 0 km/s -- channel maps of optically 
thin lines} \label{noflow} 
Figure~\ref{channel-h13co} shows channel maps of H$^{13}$CO$^+$ and
C$^{34}$S emission between +1 and --6 km s$^{-1}$. The distribution of
bulk emission of the cloud between --2 and --5 km s$^{-1}$ is similar
in both tracers and follows the typical NS-ridge of the DR21 filament.
Some emission peaks are stronger in H$^{13}$CO$^+$ than in C$^{34}$S (and
vice versa) due to the different density and temperature regimes. An
example is seen at velocity --4 km s$^{-1}$ where C$^{34}$S peaks
prominently at the DR21(OH) clump while H$^{13}$CO$^+$ emits more strongly 
in the northern part of the filament.

\begin{figure}[ht]
\includegraphics[angle=-90,width=85mm]{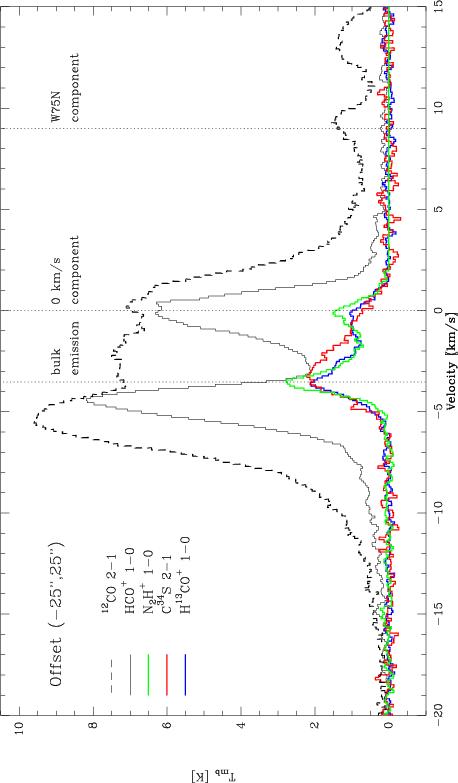}
\caption [] {Spectra of several line tracers at the position --25$''$,25$''$. 
  The $^{12}$CO line was reduced by a factor 3.5 and for the N$_2$H$^+$ line, 
  we only show the 101-012 component of the hyperfine structure (to avoid 
  confusion). The different emission components are indicated with short dashed 
  lines.} 
\label{no-outflow}
\end{figure}

\begin{figure}[ht]
\includegraphics[angle=0,width=70mm]{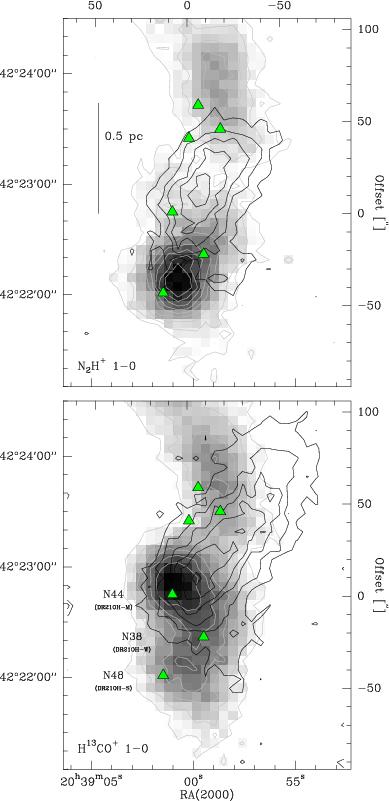}
\caption [] {Maps of line integrated N$_2$H$^+$ 1$\to$0 (top) and 
  H$^{13}$CO$^+$ 1$\to$0 (bottom) emission of the DR21(OH) region. For
  N$_2$H$^+$, we integrate over all hyperfine structure components,  
  the greyscale covering the velocity range --6 to --3 km s$^{-1}$ 
  going from 0.8 to 6.7 K km s$^{-1}$ and the black contours covering the
  range --1 to 1 km s$^{-1}$ and go from 0.32 to 1.93 in steps of 0.32
  K km s$^{-1}$. For H$^{13}$CO$^+$ the same velocity range --6 to --3
  km s$^{-1}$ represents grey scale (0.64 to 3.8 K km s$^{-1}$) and
  the black contours the velocity range --1 to 1 km s$^{-1}$ with
  contours from 0.67 to 6.6 by 0.54 K km s$^{-1}$). Green triangles
  mark mm-continuum sources, following the notation of Motte et
  al. (\cite{motte2007}).}
\label{chan-comp}
\end{figure}

At velocities larger than --2 km s$^{-1}$, we identify an extended
emission component between --1 and $0.5\,\mathrm{km}\,\mathrm{s}^{-1}$
that appears prominent in H$^{13}$CO$^+$ (and N$_2$H$^+$ not shown
here) but not in C$^{34}$S (and H$_2$CO also not shown here) northwest
of DR21(OH).  In $^{12}$CO and $^{13}$CO 2$\to$1, this feature is
discernible as well, but largely blended with other emission
components. This sub-filament is the eastern end of the large-scale
filament F3 that was observed in lower-angular resolution $^{13}$CO
1$\to$0 channel maps (Fig.~\ref{fcrao-channel}). It was also seen in
isotopomeric CO 3$\to$2 data (Vall\'ee \& Fiege ~\cite{vallee2006})
and interpreted by the authors as being caused by outflow emission from
DR21OH(M). An outflow is indeed detected at this position (see 
Sect.~\ref{outflows}) but the emission in the velocity range
$\sim$--1 to 1 km s$^{-1}$ is clearly due to an individual cloud
fragment. Spectra towards a position within this fragment are shown in
Fig.~\ref{no-outflow}. An individual Gaussian line is seen at 0 km
s$^{-1}$ for all optically thin lines (at this position rather close
to DR21OH(M) also for C$^{34}$S). It becomes obvious that the
$^{12}$CO 2$\to$1 line -- even at an angular resolution of 11$''$ --
is not useful for separating individual components since it is too
sensitive to low-density emission.  Figure~\ref{chan-comp} superimposes 
N$_2$H$^+$ and H$^{13}$CO$^+$ emission in the two major
velocity ranges, the bulk emission between --6 to --3 km s$^{-1}$ in
grey scale and the fragment emission between --1 and 1 km s$^{-1}$ as
overlaying contours.  The emission in H$^{13}$CO$^+$ for both velocity
ranges peaks at the position of DR21OH-(M). A secondary weaker peak is
found $\sim$50$''$ further south, close to N48 (or
DR21OH-(S))\footnote{In this paper, we follow the notation of
mm-continuum clumps given in Motte et al. \cite{motte2007}. This
source list is more complete than the one from Chandler et
al. \cite{chandler1993}. However, for clarity, we still give the
former names DR21OH(Main, South, West) on occasion.}, where N$_2$H$^+$
between --6 and --3 km s$^{-1}$ has its peak emission. This supports
the 'hot core' scenario for DR21-OH(M) because the gas there is warmer
and chemically richer than in the dense, cold core of DR21OH-(S). This
core, however, is embedded in a warmer envelope traced by
H$^{13}$CO$^+$.  Even more interesting is the 0 km s$^{-1}$
component. The emission distribution in H$^{13}$CO$^+$ and N$_2$H$^+$
suggests that this fragment 'falls' on the densest part of the
DR21OH(M) clump. The N$_2$H$^+$ peak is more distinct and smaller in
extent and shifted with respect to H$^{13}$CO$^+$, again indicating
that H$^{13}$CO$^+$ emission arises from the warm envelope.

\subsection{Ouflows} \label{outflows} 
\begin{figure}[ht]
\includegraphics[angle=0,width=75mm]{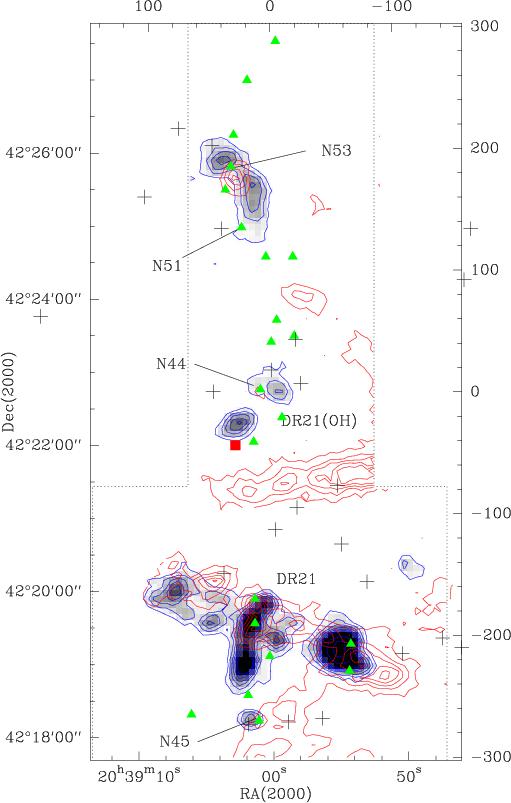}
\caption [] {Map of line integrated $^{12}$CO 2$\to$1
  emission in the DR21 filament. Blue contours indicate a velocity
  range of --30 to --17 km s$^{-1}$ (levels 12 to 60 by 8 K km
  s$^{-1}$ and then 100, 140 K km s$^{-1}$), red contours a range of 17
  to 30 km s$^{-1}$ (levels 6 to 36 by 6 K km s$^{-1}$). Green
  triangles indicate mm-continuum sources from Motte et
  al. (\cite{motte2007}), black crosses mark H$_2$ jets and knots
  discovered by Davis et al. (\cite{davis2007}), and the red square south
  of DR21-OH indicates the 'Extremely Red Object' ERO1 detected by
  Marston et al. (\cite{marston2004}).}
\label{out}
\end{figure}

Figure~\ref{out} shows in more detail the outflow features detected in
Fig.~\ref{pv}. From the position-velocity plot and positionally averaged $^{12}$CO
spectra, we can clearly define the {\sl blue} emission range between 
--30 and --17 km s$^{-1}$. The redshifted emission is
more difficult to discern because at higher positive velocities, a
part of the line emission in the southern part of the DR21 filament is
due to the 'Great Cygnus rift' (see Sect.~\ref{maps}). However, the
velocity range $\sim$17 to $\sim$30 km s$^{-1}$ characterizes well the
red wing.

Several blue outflow sources are clearly recognized. These are all
associated with mm-continuum sources, i.e. N53, N44, and N45. Source
N51 shows broad line wings in $^{12}$CO and HCO$^+$
(Fig.~\ref{spectra-plot}), which is most likely only the south extention of
the N53 outflow. Southeast of N44 and at offsets --100$''$,--140$''$
are clearly defined blue sources that have no red counterparts.
Sources N53, N48, and N44 show outflow emission in SiO 2$\to$1 (Motte
et al. \cite{motte2007}) with a prominent red wing for N53 and a blue
wing for N44. They are classified as massive infrared-quiet
protostellar cores (Motte et al.~\cite{motte2007}) with masses of 85,
197, and 446 M$_\odot$, respectively.  N45 exhibits no SiO emission
but is associated with an H$_2$ jet (source A 3-1 in Davis et
al. \cite{davis2007}), which is probably driven by a low- or
intermediate-mass YSO (all H$_2$-jets are marked with crosses in the
plot). N44 is the known outflow of DR21(OH), seen in CS 5$\to$4
emission (Richardson et al. \cite{richardson1994}) and indirectly
traced by the spatial and kinematic distribution of maser sources
(Plambeck \& Menten \cite{plambeck1990}).  However, no H$_2$ emission
was detected (Davis et al. \cite{davis2007}).

\begin{figure}[ht]
\includegraphics[angle=0,width=80mm]{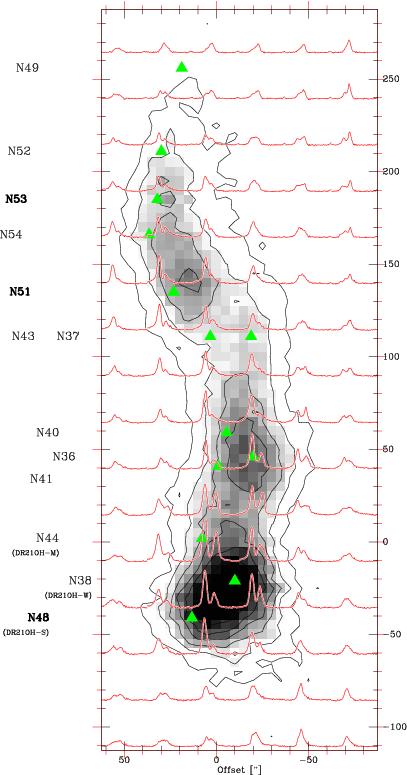}
\caption [] {Greyscale map of line integrated N$_2$H$^+$
  1$\to$0 emission of the DR21 filament with spectra of HCO$^+$
  overlaid in red. The spectra cover a velocity range of --10 to 10 km
  s$^{-1}$ and a temperature range of --1 to 12 K.  Green triangles
  indicate mm-continuum sources from Motte et al. (\cite{motte2007})
  and are named accordingly. N44, N38, and N48 coincide with the 
  mm-sources  DR21OH-M(Main),W(West), and S(South), respectively, named by 
  Chandler et al. (\cite{chandler1993}). The 0$''$,0$''$ position is DR21OH(M)
  (see Sect.~\ref{obs}). Spectra of the sources in bold (N53, N51, and N48) 
  are displayed in Fig.~\ref{spectra-plot}.}
\label{spec-over}
\end{figure}

\begin{figure}[ht]
\includegraphics[angle=0,width=60mm]{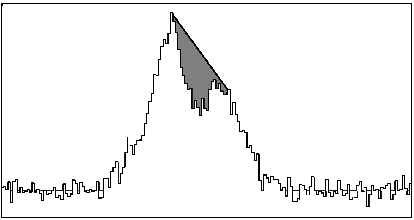} 
\caption [] {The grey area, divided by the total line-integrated intensity 
over all velocities (the 'normalized line intensity'), is shown in
Fig.~\ref{sa}. Note that only spectra with a real infall line profile
(T(blue)$>$T(red)) were considered.}
\label{norm}
\end{figure}

\begin{figure}[ht]
\includegraphics[angle=0,width=60mm]{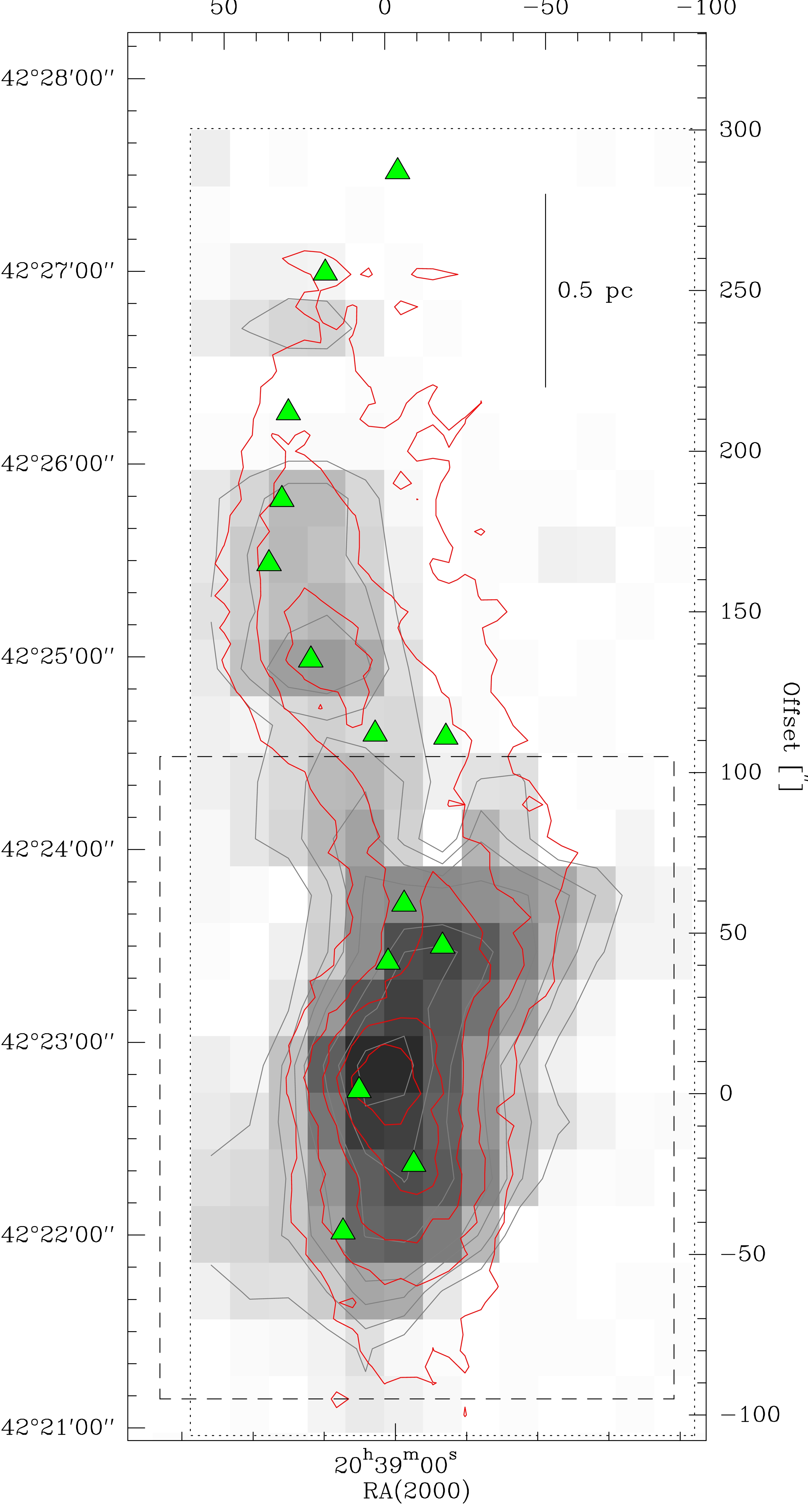}
\caption [] {Greyscale map of the ratio of 
  self-absorption to line-integrated HCO$^+$ 1$\to$0 emission in the
  DR21 filament. Overlaid on that are red contours of H$^{13}$CO$^+$
  1$\to$0 emission (contours go from 3 to 13 K km s$^{-1}$ by 1.5 K km
  s$^{-1}$).  Green triangles indicate mm-continuum sources. The
  rectangle indicates the zoom region displayed in Fig.~\ref{chan-comp}.}
\label{sa}
\end{figure}

This source is well studied (e.g. Garden et al. \cite{garden1991a},
\cite{garden1991b}, \cite{garden1992}), so we do not go into
details here.  In our map, we see also the classical east-west
orientation of the flow with some north-south features (e.g. component
'C' in Garden et al.). Among the other sources, only N53 and N44 show
red wing emission. It is not clear what the emission feature running
east-west at offsets --60$''$ in declination is. In Fig.~\ref{out}, it
shows up in HCO$^+$ emission as a typical outflow feature. But it has
no blue counterpart and may well be associated with the Cygnus Rift.

\begin{figure}[ht]
\includegraphics[angle=0,width=70mm]{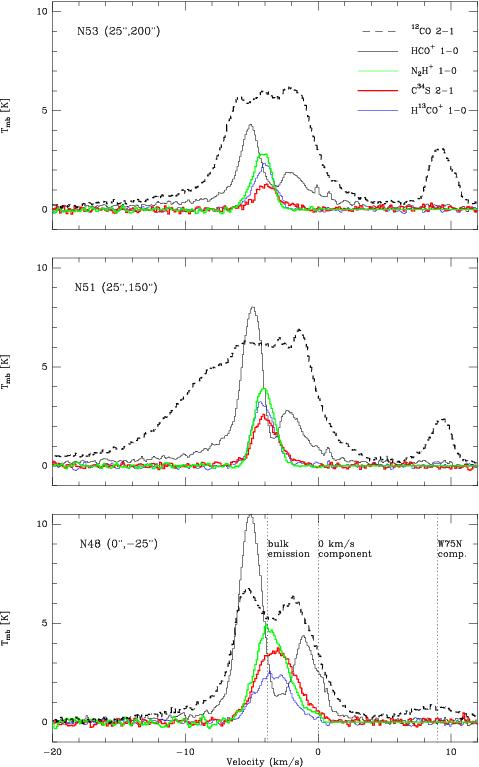}
\caption [] {Spectra of optically thick ($^{12}$CO 2$\to$1 and HCO$^+$ 1$\to$0) 
  and thin (N$_2$H$^+$ 1$\to$0, C$^{34}$S 2$\to$1,and H$^{13}$CO$^+$)
  spectra from selected positions in the DR21 filament. The $^{12}$CO
  intensity was reduced by a factor of 4. To avoid confusion, the
  lower frequency HFS components of the N$_2$H$^+$ line were
  suppressed. }
\label{spectra-plot}
\end{figure}

\section{Analysis} \label{analysis} 
This section is devoted to an analysis of the physical properties of
the DR21 filament.  We, however, exclude the area around DR21 since
this region is perturbed by outflow emission and line confusion. We
focus instead on the northern part of the DR21 filament, including the
region of DR21(OH).

\subsection{Physical properties of the sub-filaments}   \label{subfilaments} 

Our molecular line observations show that the DR21 filament is
connected to several sub-filaments seen on large and small scales. In
the $^{13}$CO 1$\to$0 channel maps (Fig.~\ref{fcrao-channel}), it
becomes obvious that the most prominent filament F3 covers a velocity
range of $\sim$3 km s$^{-1}$ (between --4 km s$^{-1}$ and --1 km
s$^{-1}$) and has a projected length of $\sim$7 pc (the width is
$\sim$1.5 pc). The channel maps show that the top of the column lies
at higher velocities, indicating that this part of the filament is
tilted away from the observer.  Assuming an average angle to the
line-of-sight of 57.3$^\circ$ (in the case of a random distribution of
orientation angles, the average angle is equal to this value), this
implies a dynamic lifetime of $\sim$1.5$\times$10$^6$ yr. The mass of
the filament (for more details see Appendix A about the mass
determination from $^{13}$CO data) is 2600 M$_\odot$ and the average
density is 690 cm$^{-3}$. This sub-filament indeed corresponds to 'clump 7'
in the DR21 region, identified in the $^{13}$CO 2$\to$1 survey by
Schneider et al. (\cite{schneider2006}).

The higher angular resolution IRAM observations resolve the detailed
structure and show how this sub-filament connects to the DR21(OH)
clump within the filament (see Sect.~\ref{noflow}). The mass,
determined from the line integrated H$^{13}$CO$^+$ intensity (see
Appendix A), is 1320 M$_\odot$ with an average density of
8.1$\times$10$^4$ cm$^{-3}$. This mass estimate, however, is rather
uncertain because of the uncertain [HCO$^+$]/[H$_2$] abundance
ratio. From the channel maps (Fig.~\ref{channel-h13co}), we see that
the filament extends up to $\sim$ 0.5 km s$^{-1}$ and therefore has a
total radial velocity range of 4.5 km s$^{-1}$ at a length of $\sim$1
pc. If we assume that it connects to the large filament F3, the lower
velocity levels are equally around --4 km s$^{-1}$, thus implying a
total velocity range of 4.5 km s$^{-1}$.  Again, assuming an average
angle to the line-of-sight of 57.3$^\circ$, this implies a very short
dynamic lifetime of $\sim$1.4$\times$10$^5$ yr.

In addition, we note that in the high spatial resolution IRAM data,
this sub-filament shows a position angle of $\sim$45$^\circ$ in
projection to the sky, while on a larger scale F3 is mostly east-west
oriented (PA $\sim$ 100$^\circ$). This bend, which could be related to
the deep gravitational well of the DR21(OH) clump, may also be present
along the line-of-sight, explaining the apparent acceleration from 3
to 4.5 km s$^{-1}$ along the sub-filament between large and small
scales. Assuming a constant velocity of 4.5 km s$^{-1}$ (the largest
observed range of velocity) along the whole F3 sub-filament, the angle
of the sub-filament to the line-of-sight would be 48$^\circ$ before
the bend and 90$^\circ$ after the bend, close to DR21(OH)
(i.e. streaming away from the observer along the line-of-sight). This
3D view of F3 seems to be corroborated by the largest velocities in
the IRAM data being found closest to DR21(OH). The total length of the
F3 sub-filament would then be at least 9.4~pc (7 pc divided by
sin(48$^\circ$)) with no information about the length along the
line-of-sight after the bend, leading to a total dynamic timescale of
$\sim$2.1$\times$10$^6$ yr, and an input mass rate of
$\sim$1.9$\times$10$^{-3}$ ~M$_\odot$ yr$^{-1}$ (total mass of F3 on
large and small scales divided by the timescale).

\subsection{Infall signatures across the filament }   \label{infall} 
Figure~\ref{spec-over} shows an overlay of HCO$^+$ 1$\to$0 spectra
(red) on a greyscale plot of line integrated N$_2$H$^+$ 1$\to$0
emission (both at an angular resolution of $\sim$30$''$). It is 
obvious that HCO$^+$ displays {\sl along the whole filament} the
'typical' profile of an optically thick spectral line which is indicative 
of infalling gas, i.e. a blue-shifted line wing that is stronger than the
red wing (Myers et al. \cite{myers1996}). This self-absorption was
already visible in $^{12}$CO emission (Fig.~\ref{pv}) but becomes more
pronounced using the HCO$^+$ line.  In order to quantify this infall
signature, we determined the normalized integrated intensity (see
Fig. ~\ref{norm} between the blue and red line components (with
T(blue)$>$T(red)), i.e. the 'amount' of self-absorption, and plotted
this quantity in Fig.~\ref{sa}. It is remarkable that self-absorption is seen
everywhere in the filament on a size scale of $\sim$10$'$=5 pc with
smooth transitions and not only locally at the position of
mm-continuum sources.  However, spectra with strongest self-absorption
are found at the position of N44 (DR21OH(M)).

To exclude the possibility that we observe several line components, we
display in Fig.~\ref{spectra-plot} examples of various lines at three
selected positions (that are modeled in Sect.~\ref{simline}). First,
the spectra show that $^{12}$CO is not the most reliable infall tracer
since it is ubiquitous and the infall profile is most clearly seen in
HCO$^+$.  The figure also clearly proves that not two different cloud
components are observed since all optically thin lines (C$^{34}$S,
N$_2$H$^+$, H$^{13}$CO$^+$) peak in the emission gap of HCO$^+$.
Apart from the known line components at --3 km s$^{-1}$ (bulk emision
of the DR21 filament) and +9 km s$^{-1}$ (W75N, only visible in
$^{12}$CO 2$\to$1), the 0 km s$^{-1}$ component (see
Sect.~\ref{noflow}) is best traced in N$_2$H$^+$ emission.

Since it is assured that we observe a true infall signature across the
DR21 filament, we determined the infall velocity v$_{in}$ to first
order using the method described in Myers et al. (\cite{myers1996}).
\begin{eqnarray}
{\rm v}_{in} & = & \frac{\sigma^2} {{\rm v}_{red}-{\rm v}_{blue}} \ln  \frac{1+e(T_{BD}/T_{D})}{1+e(T_{RD}/T_{D})}
\label{inf}
\end{eqnarray}

\begin{figure*}[ht]
\includegraphics[angle=-90,width=140mm]{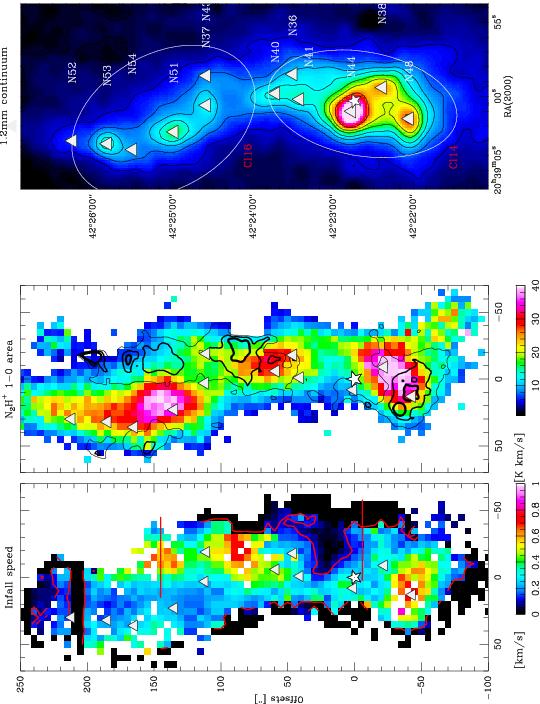}
\caption [] {{\bf Left:} Color plot of infall velocity. 
  The red contour line indicates the thermal sound speed of 0.2 km
  s$^{-1}$. White triangles mark mm-continuum sources. 
  The two red lines indicate the cuts where we model spectra 
  (Sect.~\ref{simline}). The cuts were selected to cross 
  the center of each ellipse. {\bf Middle:}
  Contours of infall velocity (levels are 0.3, 0.5, 0.7 km s$^{-1}$)
  overlaid on a plot of line-integrated N$_2$H$^+$ intensity
  determined from a Gaussian line fit to the --3 km s$^{-1}$
  component. It becomes obvious that the highest infall velocities 
  are offset from density maxima.
  {\bf Right:} Mm-continuum emission map taken 
  from Motte et al (\cite{motte2007}), where the two ellipses indicate 
  the clumps considered for modeling and calculating infall 
  properties. Note that all plots cover exactly the same area and 
  have the same scales. }
\label{infall-speed}
\end{figure*}

\noindent Here, $\sigma$ is the H$^{13}$CO$^+$ velocity dispersion,
$T_{RD}$, v$_{red}$ and $T_{BD}$, v$_{blue}$ the temperature and the
velocity of the red and blue part of the absorption profile,
respectively, e=2.71828 the Eulerian number, and T$_{D}$ is the
brightness temperature of the dip.  In all calculations, we excluded
the area containing DR21 ($\Delta\delta <$--100$''$).  The velocity
dispersion is determined from the FWHM line width of the optically
thin H$^{13}$CO$^+$ line
($\sigma$=FWHM(H$^{13}$CO$^+$)/$\sqrt{8\,\ln(2)}$).  We disentangle
the optically thin emission of the molecular ridge ($\sim$--3 km
s$^{-1}$) from the infalling sub-filament at 0 km s$^{-1}$ with a
double Gaussian fit that precisely determines $\sigma$ \ of the bulk
emission to be --3 km s$^{-1}$.  For the optically thick HCO$^{+}$
line, we did not perform such a double Gaussian fit since the --3 km
s$^{-1}$ component, showing the infall signature, dominates strongly
over the single Gaussian 0 km s$^{-1}$ component (this means that the
subfilament does {\sl not} show an infall signature).  We determined
the temperature and the velocity of the red ($T_{RD}$, v$_{red}$) and
blue ($T_{BD}$, v$_{blue}$) part of the absorption profile by fitting
a Gaussian to each wing.  We only computed the infall speed in a
region where the emission of the second component does not contribute
significantly and confuse the absorption profile.  We note that we
consider each point in the map as a single two-layer model (as
outlined in Myers et al. \cite{myers1996}).

Figure~\ref{infall-speed} shows the results of this procedure.
Supersonic infall velocities, v$_{in}>$0.2 km s$^{-1}$ for gas of a
temperature of 10 K (typical value found), are detected within the red
contour line, i.e. {\sl covering a large part of the
filament}. Interestingly, the highest values of v$_{in}$
($\sim$0.6--0.8 km s$^{-1}$) are not detected exactly towards the
column density maxima (indicated by N$_2$H$^+$ and mm-continuum
emission in the right panels of Fig.~\ref{infall-speed}), but,
instead, show some significant spatial offsets ranging from
20$^{\prime\prime}$ to 30$^{\prime\prime}$ (i.e. of the order of
0.2~pc) in projection. In the center region (DR21OH(M) and N40) and in
the south of the filament (N38 and N48), high values of v$_{in}$ are
mainly due to the increasing optical depth $\tau$ of the HCO$^+$ line
since the infall velocity is determined by the ratio $T_{BD}$/$T_{RD}$
(eq.~\ref{inf}), which increases with $\tau$.  In the north, higher
values of v$_{in}$ are offset by $\sim$20$''$ to the dense clump
containing N51 to N54 and are mainly related to the larger linewidths
of the N$_2$H$^+$ line (see Fig.~\ref{fit-h13co}, showing the same
increase in linewidth for H$^{13}$CO$^+$ displaced with respect to the
dense clump).

Please note that this behaviour does {\bf not} involve a gradual {\sl
increase} in infall speed since we only determine the infall velocity
at each point of the map assuming local infall at each point.  From
this figure, we can not extract the radial profile of the infall speed
within the whole filament. For that, we need a more realistic
radiative transfer model, by considering the density and temperature
structure in the dense clumps of the filament as we present in the
next section.

\subsection {Radiative transport modelling with Simline} \label{simline} 
We used the 1D non-LTE radiative transfer code \textit{Simline}
(Ossenkopf et al. \cite{ossenkopf2001}) to simultaneously model the
observed optically thin H$^{13}$CO$^+$ and optically thick HCO$^+$
lines of the two clumps indicated on the right panel of
Fig.~\ref{infall-speed}.  These were extracted from the mm-continuum
survey obtained by Motte et al. (\cite{motte2007}) and characterize
well the two major regions of infall. The southern
\textsl{Clump-14} is associated with DR21(OH) and contains the
cores N36, N38, N41, N44, and N48. In the northern part of the
filament, \textsl{Clump-16} contains the cores N37, N43, N51, N53, and
N54. The advantage of using mm-continuum data is that we have a rather
reliable mass estimate than can be used as input to the model.  Our
ultimate goal is to derive the infall velocity and the level of
depletion by constraining a simple physical model for these clumps.

\subsubsection{Input parameters and assumptions} 
For both clumps, we used the Gaussian parameters (flux and {\it
FWHM}) determined by Motte et al. (2007) and assumed the more
realistic $\rho(r)\propto r^{-2}$ density profile to estimate the
$90\%$ mass and corresponding radius. We also recalculated
the mass down to the level of 10\% of the peak contour {\sl Clump-16}
taking a mass-averaged temperature of 20~K (rather than the median
temperature of Cygnus~X clumps, 15~K, used by Motte et al. 2007), which
is in closer agreement with NH$_3$ measurements of this clump (Wienen 
2008). The model self-consistently takes the mass corresponding
to the mass-averaged temperature with a profile $T(r) = T_{in}
\Big(\frac{r}{r_{in}} \Big)^{\beta}$, where $\beta$ is between --0.2
and --0.6 ($T_{in}$ and $r_{in}$ are the temperature and radius of the
inner layer, see below).  We adapted a two-layer geometry for the
model since runs with only one layer did not reproduce well the
observed line profiles, a result similar to that of Jakob et
al. (\cite{jakob2007}) who equally applied a two-layer model for their
low-angular resolution mid-J CO and atomic CI lines. However, they
modeled their observations without considering molecular depletion and
no or only very low infall speeds. We applied a subthermal external
layer at constant temperature with an average density of
n$\sim$10$^3$-10$^4$ cm$^{-3}$ because the self-absorption dip is very
deep.

For the determination of the turbulent line-width, we used the
optically thin line of H$^{13}$CO$^+$.  A single Gaussian fit yielded
line-widths of $\sim$2.5$\pm$0.5 km s$^{-1}$ for {\sl Clump-14} and
$\sim$1.5$\pm$0.5 km s$^{-1}$ for {\sl Clump-16}, respectively. This
line-width is a result of thermal width, internal turbulence, and
possible infall. Since both turbulence and infall broaden the line,
their parameter space must be explored carefully.

For the HCO$^+$ abundance, we take a value of {\sl X(HCO$^+$})
1-5$\times$10$^{-9}$ (e.g. Marseille et al. \cite{marseille2008}) and
adopt an isotope ratio $^{12}$C/$^{13}$C = 67 (e.g. Lucas \& Liszt 
\cite{lucas1998}). However, HCO$^+$ is known to be depleted in
cold environments and at densities higher than $\sim$10$^5$ cm$^{-3}$
(Tafalla et al. \cite{tafalla2002}). We expect this happens in {\sl
Clump-16} but to a lesser extent in {\sl Clump-14} since this region
contains a hot core. We thus introduce a {\sl threshold density}
n$_{tresh}$ in our model, with a sharp jump in the abundance profile
if n$ > {\rm n}_{tresh}$. The depletion factor is assumed to be in the
range 1-100.

Our strategy to constrain the infall speed was to vary the depletion
ratio, the abundance, and the threshold density (between the given
limits). To determine the best fit to the model, we performed a {\sl
reduced $\chi^2$}-test.

\begin{figure*}[ht]
\includegraphics[angle=0,width=160mm]{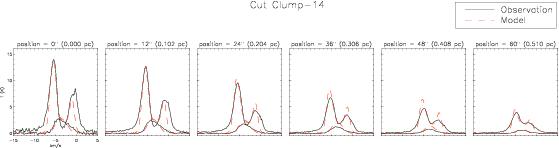}
\caption [] {Observed (black) and modeled (red) HCO$^+$ and H$^{13}$CO$^+$ 1$\to$0 spectra 
of {\sl Clump-14} (see Fig~\ref{infall-speed} for the location of the cut).}
\label{map-cl14}
\end{figure*}

\begin{figure*}[ht]
\includegraphics[angle=0,width=160mm]{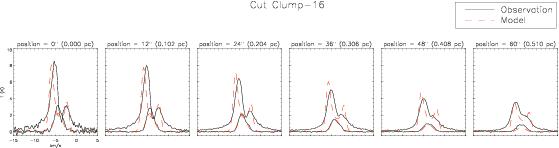}
\caption [] {The same line Fig.~\ref{map-cl14} but for {\sl Clump-16}.}
\label{map-cl16}
\end{figure*}

\begin{table}[htbp]
   \begin{tabular}{lcc}
\hline
\hline
& Clump-14 & Clump-16 \\
\hline
total mass                   & 4900 M$_{\odot}$        & 3346 M$_{\odot}$ \\
size                         & 0.52 pc                 & 0.615 pc \\
density profile ($\alpha_2$) & --8.0                   &  --8.0     \\
mass-averaged T              & 20 K                    &  15 K    \\ 
$\beta$                      & --0.6                   & --0.5    \\ 
line width                   & 2.5 km s$^{-1}$         & 1.5 km s$^{-1}$ \\ 
X(HCO$^+$)                      & 2.5$\times$10$^{-9}$    & 2.5$\times$10$^{-9}$ \\ 
$^{12}$C/$^{13}$C            & 67                      &  67       \\
depletion ratio              & 7                       &  28 \\
$\tau_{HCO^+}$               & 25                      & 26  \\ 
$\tau_{H^{13}CO^+}$          & 0.81                    & 0.99 \\ 
v$_{inf}$                    & --0.6 km s$^{-1}$       & --0.5 km s$^{-1}$  \\
\end{tabular}
\caption{Results of {\sl Simline} modelling where $\alpha_2$ is the density profile 
exponent at the transition to the subthermal layer and X(HCO$^+$ the HCO$^+$ abundance.
}
\label{simline-table}
\end{table}

\subsubsection{Results} 
Figures~\ref{map-cl14} and \ref{map-cl16} show the observed and
modeled spectra of HCO$^+$ and H$^{13}$CO$^+$ for {\sl Clump-14} and
{\sl Clump-16}, respectively. For better orientation, the two cuts are
indicated in Fig.~\ref{infall-speed}. Table~\ref{simline-table}
summarizes the explored parameter space and gives the results for the
best fitting model. We note that the spectra of both sources have
broad wings caused by outflow emission that we did not account for in
our modeling. \\

\noindent{\bf Clump-14}\\
For {\sl Clump-14}, our best-fit
model closely reproduces the spectral lines of HCO$^+$ ($\tau
\sim 25$) and H$^{13}$CO$^+$ ($\tau \sim 0.8$). 
It was necessary to use a rather large outer layer to reproduce well
the absorption dip, which extends to R $\sim$ ~0.9 pc increasing by
20\% the mass of our model.  We found that a temperature profile of
$\beta \sim -0.6$ and a mass-averaged temperature of 20 K provides a
closest fit to our data, where the discrepancy in the peak intensity
of the lines is smaller than 20\%. This also indicates that the inner
part of this clump tends to be hotter, reaching up to $\sim$100 K in
the central part.

The depletion occurs at a radius of 0.25 pc with a depletion ratio of
$\sim 7$. This may be due to the temperature because the central
region is a hot molecular core.  With this configuration of geometry
and parameters, we derive an infall speed of {\bf $\sim$ --0.6 km
s$^{-1}$}, which compares well with the values derived from the Myers
et al. method (see Sect.~\ref{infall}). \\

\noindent{\bf Clump-16}\\
The best-fit model for {\sl Clump-16} (Fig.~\ref{map-cl16}) reproduces
very well the H$^{13}$CO$^+$ line ($\tau \sim 1$) but less well the
optically thick HCO$^+$ line ($\tau \sim 26$).  Since the densities
are lower for this clump, the extended outer layer is smaller than
that of {\sl Clump-14} and reaches out to R$_{ext} \sim$0.8 pc
reproducing the 10$^3$ cm$^{-3}$ density.  As we kept the exponent
$\beta$ for the temperature profile free, we found that $\beta = -0.5$
most closely fits our data. This results in a maximum of T$\sim$60 K
in the central region. The depletion occurs at R$\sim$ 0.2 pc with a
depletion ratio of 28, implying that in the central, very dense part
of this clump, this molecule is strongly depleted.

The fit value of the infall speed is {\bf $\sim$ --0.5 km s$^{-1}$}, again
within the range obtained from the more simple Myers et
al. method. The strongest constraint on the infallvelocity is the
ratio of the left-to-right wing intensity ratio, because the position
of the plotted spectra is aligned with the optically thin line.
 
\begin{table}[ht]   \label{infall-para}
\begin{minipage}[t]{\columnwidth}
\caption{\small Infall properties for Clump 14 and 16, i.e. the two main infall regions.} 
\centering
\renewcommand{\footnoterule}{}  
\begin{tabular}{lccccc}
\hline \hline
         & R$^a$ &M$^b$       &$<$n$_{\rm H_2}>^c$ & v$ _{in}^d$  & dM/dt$^e$ \\
         & [pc]  &[M$_\odot$] &10$^5$ [cm$^{-3}$]  &[km s$^{-1}$] &[M$_\odot$ yr$^{-1}$] \\
\hline
Clump 14 & 0.52 & 4900       & 1.26               & 0.6          & 5.8 10$^{-3}$  \\
Clump 16 & 0.615 & 3346       & 0.52               & 0.5          & 2.8 10$^{-3}$   \\
\hline 
\end{tabular}
\end{minipage}
\vskip0.1cm
\noindent $^a$ Equivalent radius of infall region \\
$^b$ Mass from mm-continuum (see Table~2) \\
$^c$ average H$_2$ density, derived from the mass given in column 2 \\
$^d$ infall speed, determined from {\sl Simline} modelling (Sec.~\ref{simline})\\
$^e$ mass infall rate $\dot M \approx$  M/t = M v$_{in}$/R \\ 
\end{table}

\subsection{Global infall properties of the DR21 filament}

Table~3 summarizes the infall properties of the two main infall
regions that we defined in the {\sl Simline} modeling. It is first
important to understand that we discuss here the {\sl global} dynamic
features of the molecular {\bf clumps} on a size scale of $\sim$0.5
pc, and not the physical properties of individual {\bf cores} on a
size scale of $\sim$0.1 pc.

The large global infall speeds on a $\sim$0.5 pc size scale and the
high masses of the clumps lead to large global infall rates of $\dot
M$ of 5.8$\times$10$^{-3}$ M$_\odot$ yr$^{-1}$ and
2.8$\times$10$^{-3}$ M$_\odot$ yr$^{-1}$ for {\sl Clump-14} and {\sl
Clump-16}, respectively. However, one has to distinguish between
large-scale flow motions and local core collapse. The expression
'infall' , representing material that is falling onto a protostar or
-- more generally -- into a gravitational potential well that could
form many stars within a core that is gravitationally collapsing has
to be used with care. In Fig.~\ref{infall-speed}, we see that the
highest values of 'infall' speed do not always correspond to dense
cores, only source N48 shows a local peak for v$_{in}$. For the rest
of the DR21 filament, the highest 'infall' speeds are found in the
less dense western part of the filament and not in the dense
post-shock gas. In view of the high mass infall rates, this implies
that there is a significant contribution of matter input either by
directed flows (as would be the case for large-scale convergent flows)
or gravitational collapse of the DR21 filament itself on a large
scale. Both clumps have a low virial parameter $\alpha$\footnote{The
virial parameter is defined to be $\alpha$=M$_{vir}$/M where the virial
mass is M$_{vir}$[M$_\odot$]=699 $\sigma^2 [{\rm km/s}]^{-2}$ R [pc] and
the mass M from Table~2.} of 0.46 and 0.30 for {\sl Clump 14} and {\sl
Clump 16}, respectively, and are thus strongly gravitationally
bound. Their free-fall lifetimes t$_{ff}$\footnote{t$_{ff}$[yr]=1.37
10$^6 \Big(\frac{10^3 {\rm cm}^{-3}}{2 n({\rm H}_2)}\Big)^{0.5}$} are
0.9$\times$10$^5$ yr ({\sl Clump 14}) and 1.3$\times$10$^5$ yr ({\sl
Clump 16}), a factor 20--30 shorter than the sound crossing time,
which indicates that they are indeed in global collapse (if the simple
picture of a collapse onto a single gravitational potential is
maintained). On the other hand, the infall rates are in agreement with
the theoretical predictions of both the micro-turbulent core model
(McKee \& Tan \cite{mckee2003}) and the gravo-turbulent fragmentation model
(e.g. Klessen et al. \cite{klessen1998}, \cite{klessen2000},
\cite{klessen2001}, Padoan \& Nordlund \cite{padoan2002}, Banerjee \&
Pudritz \cite{banerjee2007}).

\subsection{The kinematic structure of the DR21 filament }   \label{kine} 
To further characterize the kinematic structure of the DR21 filament,
we determine the line velocity and velocity dispersion using
H$^{13}$CO$^+$ and N$_2$H$^+$ by fitting a Gaussian profile to the --3
km s$^{-1}$ component as described in Sect.~\ref{infall}.  For
N$_2$H$^+$, we additionally fitted the hyperfine structure pattern of
the molecule.

\subsubsection{Velocity dispersion} 
\begin{figure}[ht]
\includegraphics[angle=-90,width=88mm]{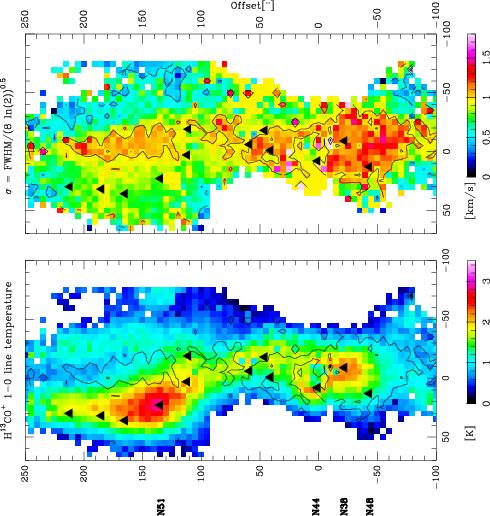}
\caption [] {Color scale maps of the results of Gaussian line fitting (temperature  
 and velocity dispersion) to the --3 km s$^{-1}$ component of
 H$^{13}$CO$^+$ 1$\to$0 along the DR21 filament (but excluding the
 region of DR21). Contours of velocity dispersion (second panel) at 0.75 and 1
 km s$^{-1}$ are overlaid on both maps. }
\label{fit-h13co}
\end{figure}

Figure~\ref{fit-h13co} reveals that the H$^{13}$CO$^+$ main beam
brightness temperature map shows a V-shaped feature in the north of
the filament. The western component is a factor 2-3 weaker than the
eastern one, which has a pronounced peak at the position of N51. This
peak is stronger than that towards N44/DR21OH(M) (the reverse is true
for the line integrated intensity, see Fig.~\ref{maps1}).

\begin{figure}[ht]
\includegraphics[angle=-90,width=88mm]{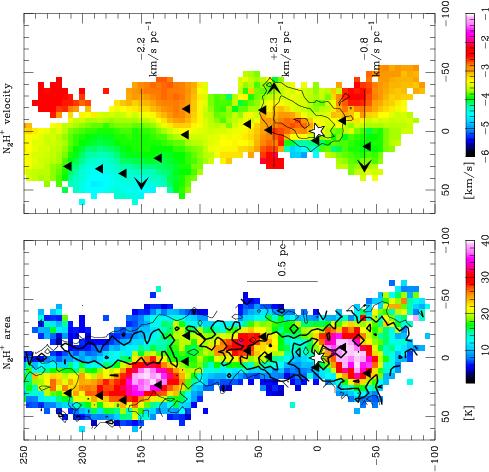}
\caption [] {Color scale maps of the results of Gaussian line fitting (line 
 integrated intensity and line velocity) to the --3 km s$^{-1}$ component of N$_2$H$^+$ 1$\to$0
 along the DR21 filament (but excluding the region of DR21). Contours
 of velocity dispersion of H$^{13}$CO$^+$ (0.75 and 1 km s$^{-1}$) are
 overlaid on the area map. Contour lines of the line integrated (--1 to 1 km s$^{-1}$) 
 emission of the 0 km s$^{-1}$ component are overlaid on the velocity map (levels 
 1, 2, 3 K km s$^{-1}$). }
\label{fit-n2h}
\end{figure}

Most remarkable, however, is the velocity dispersion $\sigma$
(determined from the FWHM of the line with $\sigma$=FWHM/($8
\ln(2))^{0.5}$) seen in H$^{13}$CO$^+$ as well as in N$_2$H$^+$ (not
shown here). The velocity dispersion {\sl increases} towards the
geometrical center of the filament from sonic levels to values of
$\sim$1.3 km s$^{-1}$. In the northern part of the filament
(declination offset$>$0$''$), the gradient is stronger than in the
southern part (south of N44/DR21OH-(M)) where we see a more
homogeneous distribution of $\sigma$.  Interestingly, in the northern
part of the region highest $\sigma$-values are found {\sl between} the
two vertical features of H$^{13}$CO$^+$ emission (with a decrease of
$\sigma$ from $\sim$1 km s$^{-1}$ to $\sim$0.8 km s$^{-1}$ in the
densest region in the east). This becomes even more obvious in
Fig.~\ref{fit-n2h} where we plot contours of $\sigma$ (from
H$^{13}$CO$^+$) over a map of N$_2$H$^+$ line integrated intensity
(proportional to the column density).  Thus, there is a clear offset
between high $\sigma$ and high column densities. This scenario can be
explained if it is assumed that clumps are produced by turbulent flows
because in this case, the densest gas has been shocked and slowed down
so that the largest velocity dispersions do not occur in the densest
regions but in the outskirts (Klessen et al.~\cite{klessen2005},
Vazquez-Semadeni et al.~\cite{vaz2008}, Federrath et
al.~\cite{fed2009}).  We return to this point in the discussion.  The
southern part (offset DEC$<$0$''$) seems to behave differently. The
N$_2$H$^+$ column density and H$^{13}$CO$^+$ line temperature are
highest in the core region ($\sim$0.5 pc diameter) of N38, while the
velocity dispersion remains on a rather constant, high-level of around
1.2 km s$^{-1}$ in an area of $\sim$1 pc diameter. However, a
projection effect could explain the difference.: it is possible that a
similar offset between velocity dispersion and highest density region
is seen face-on.

\subsubsection{Velocity field} 
Figure~\ref{fit-n2h} shows the complex velocity field of the DR21
filament determined from N$_2$H$^+$ (the H$^{13}$CO$^+$ map not shown
here looks very similar). Three main, basically horizontal, projected
velocity gradients are observed.  While the northern and southern part
show a decrease of velocity from west to east (--2.2 and --0.8 km
s$^{-1}$ pc$^{-1}$), the middle part has a positive value of +2.3 km
s$^{-1}$ pc$^{-1}$. Interestingly, we observe in this region a
'turnover' in the velocity gradient that corresponds to the material
of the 0 km s$^{-1}$ component 'falling' onto the bulk of the filament
(see Fig.~\ref{fit-n2h}).  It is clear that the velocity pattern
cannot be explained with a single rotation of the filament along a
vertical axis.  An alternative explanation would again be that at
least a part of the observed motion is due to convergent flows (see
Sect.~\ref{discuss}) and/or rotation of the individual parts of the
filament.

\begin{figure*}[ht]
\includegraphics[angle=0,width=0.5\linewidth]{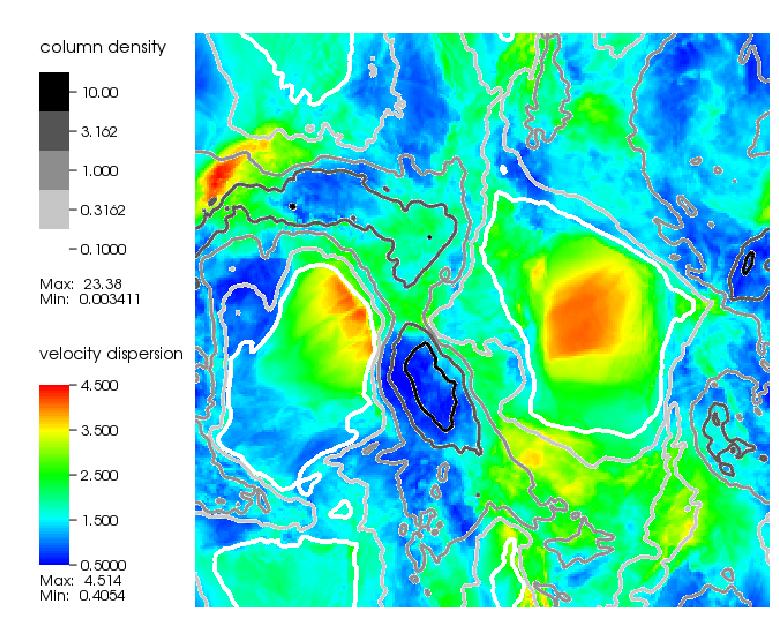}
\caption [] {Results of a hydrodynamic turbulence simulation by Federrath
et al. (\cite{fed2009}), considering only compressive (curl-free) forcing.
The Mach numbers reached in this simulation are comparable to the typical
Mach numbers observed in the DR21 filament. The color-coded velocity dispersion
is in units of the sound speed, while contours of column density are overlaid 
on it. The simulation was performed with the grid code FLASH3 and used $1024^3$
computational elements. This figure can be directly compared to 
Fig.~\ref{infall-speed} (middle panel), also showing a prominent offset between
the maxima of velocities and column densities.}
\label{model1}
\end{figure*}

\section{Discussion} \label{discuss} 

\begin{figure*}[ht]
\includegraphics[angle=0,width=80mm]{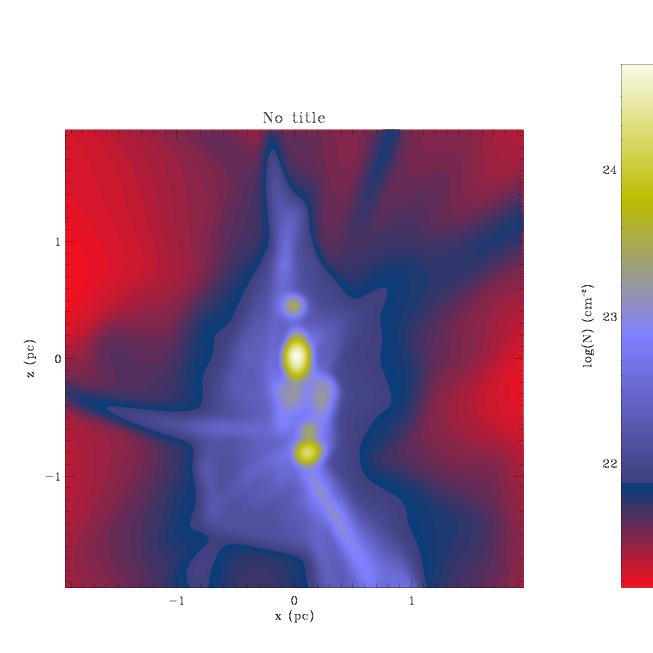}
\includegraphics[angle=0,width=80mm]{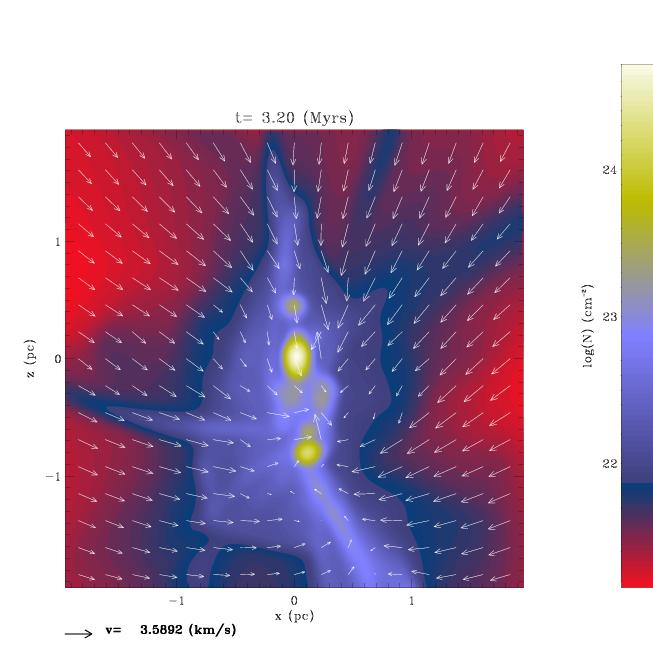}
\caption [] {In the left-hand plot, we show the column density distribution from an 
MHD code (Hennebelle et al., Teyssier \cite{teyssier2002}, Fromang et al. \cite{fromang2006}). In
the right plot, the projected velocity field is overlaid on the column
density distribution.}
\label{model2}
\end{figure*}

The dynamic structure of the DR21 filament exhibits several remarkable features: 

\begin{itemize} 
\item Several sub-filaments are connected to the DR21 filament, the most 
massive one, F3, being clearly linked to the DR21(OH) clump.
\item{} Global infall signatures are seen all across the filament (on a 
length scale of a few pc). 
\item The highest values of the velocity dispersion are found in a vertical 
projected region of low column density gas, offset from the dense clumps seen in N$_2$H$^+$. 
\item Three horizontal (west-east) velocity gradients (--0.8, +2.2, +2.3 
km s$^{-1}$ pc$^{-1}$) exist in the filament. 
\end{itemize} 

All of these findings point towards a very dynamic nature of the DR21
filament in which motions are not randomly distributed but show clear
signs of systematic dynamics. These features all point to local
converging flows being the driving source of the whole filament, with
an increase in density at the stagnation points of the flows
(Sect.~\ref{flow}). In addition, the large-scale infall signatures
observed all over the filament point towards a scenario in which {\sl
gravity} is the dominant driving force on the scale of the filament
(Sects.~\ref{collapse} and ~\ref{mhd}).  On the other hand, the origin
of the large-scale flows is difficult to constrain. It is possible
that the large-scale dynamics are a remains of the formation process
of the molecular complex, but may also originate either from
self-gravity on larger scales, or large-scale turbulence injection
(Sects.~\ref{numbers} and ~\ref{largescale}).

\begin{figure*}[ht]
\includegraphics[angle=0,width=75mm]{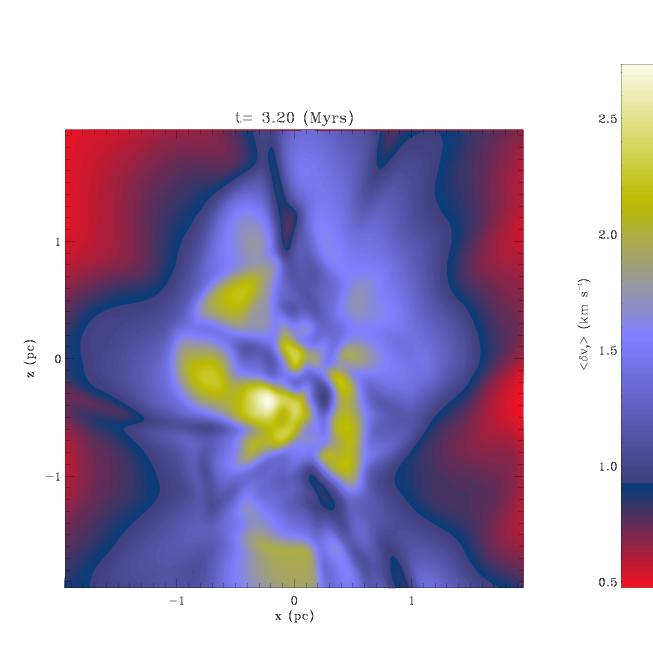}
\includegraphics[angle=0,width=75mm]{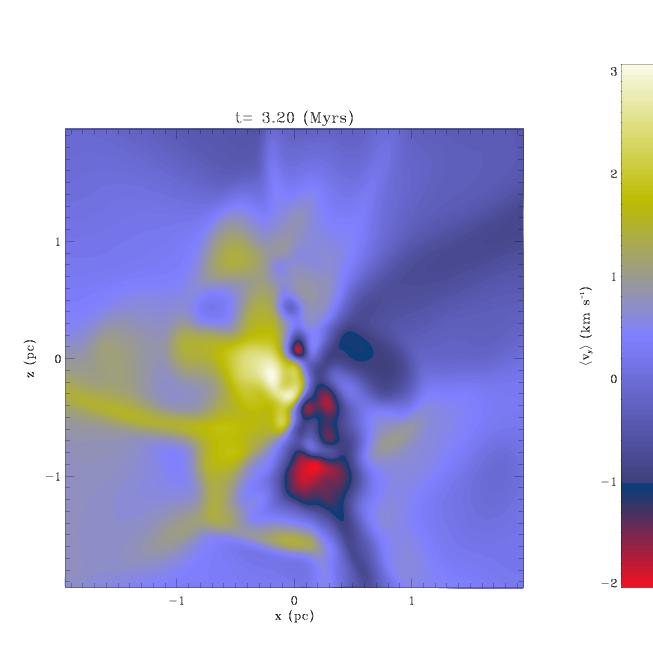}
\includegraphics[angle=0,width=75mm]{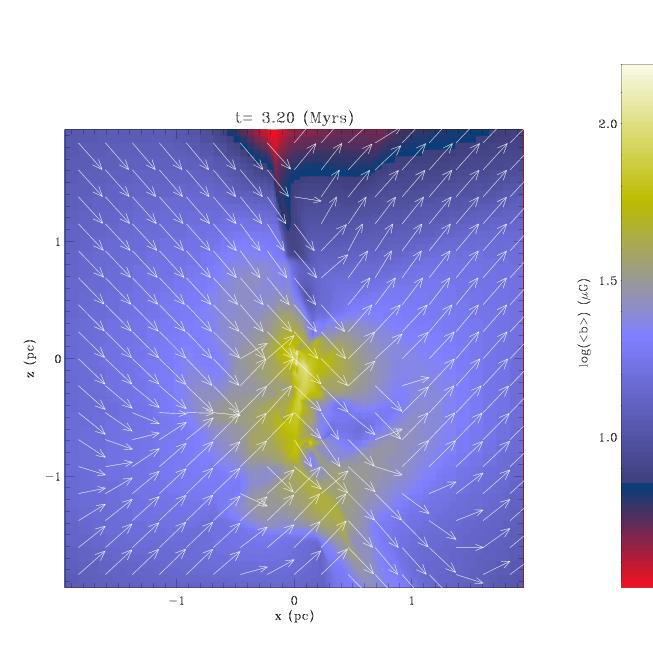}
\caption [] {Results from the same MHD turbulence model (Hennebelle et al., 
Teyssier \cite{teyssier2002}, Fromang et al.  \cite{fromang2006}) 
shown in Fig.~20. {\bf Top:} Velocity dispersion (left) and velocity 
field (right). {\bf Bottom}: Magnetic field with field vectors overlaid on it.}
\label{model3}
\end{figure*}

\subsection{Organized motions: signposts of convergent flows} \label{flow} 

The clouds of the whole Cygnus North region are very similar in their
physical properties (morphology, density etc.) to clouds that were
created by colliding flow scenarios in models (e.g. the final stage of
the Gf2 model from Heitsch \& Hartmann \cite{heitschhart2008}). The
DR21 filament is part of a complex network of interconnected
filamentary structures (Sec.~\ref{fcrao}) and could thus indeed be the
result of a convergent, turbulent flow.  In particular, the offset of
the infall speed from the densest regions is consistent with a
scenario where supersonic flows shock at their stagnation points are
offset from the dense postshock gas on both sides of the filament.  In
this case, high-density clumps/cores are built up at the stagnation
points of the colliding flow and would explain the distribution of the
dense mm-cores along the DR21 filament. It suggests that the cores are
dynamically distributed through the region before they evolve into
intermediate or high-mass star-forming clusters. From the mm-continuum
survey of Motte et al. (\cite{motte2007}), we determined that
approximately 15\% of the mass of the filament is contained in dense
cores, in accordance with the predictions of Heitsch \& Hartmann
(\cite{heitschhart2008}).
 
Another observational results that agrees  with supersonic
isothermal simulations is the velocity field and the distribution of
the velocity dispersion ($\sigma$).  We observe different velocity
gradients along the filament. Figure~\ref{fit-n2h} shows in the
northern and southern part of the filament a negative gradient in
the WE-direction whereas at the position of DR21(OH) (where the
subfilament falls onto the densest region), this gradient is
reversed. It is unlikely that the gradients are only caused by
rotation (see, e.g. Goodman et al. \cite{goodman1993} for typical
observational signatures of uniform rotation). Moreover, they can be
explained in the turbulent colliding flow view, where clumps are
sheared and compressed due to the flow motions and retain the signature
of the external flow that formed them (Ballesteros-Paredes et
al. \cite{ball1999}, Vazquez-Semadeni et al. \cite{vaz2008}) so that
only parts of the filaments are rotating.

The shock front created by the colliding flows creates an inner dense
post-shock region and an outer lower-density region of supersonic
inflow. Cores are then formed by 'gravoturbulent fragmentation' (Klessen et
al. \cite{klessen2005}, Hennebelle \& Chabrier
\cite{hennebelle2008c}), which characterizes a hierarchy of smaller and
smaller entities. The local collapse of a core occurs when
self-gravity overcomes the gas pressure (including the turbulent
contribution to pressure). The whole process leads to a higher
velocity dispersion {\sl outside} the clump/core than in the more
quiescent post-shocked interior (Klessen et al. \cite{klessen2005},
Gomez et al. \cite{gomez2007}, Vazquez-Semadeni et
al. \cite{vaz2008}), which is precisely what we observe in the DR21
filament. Figure 1 in Klessen et al. (\cite{klessen2005}) illustrates
this scenario in a non-magnetic SPH (smoothed particle hydrodynamics)
model.  Here, we provide evidence of the same scenario in
Fig.~\ref{model1}, where we show a map of the velocity dispersion
obtained in a driven turbulence simulation by Federrath et
al.~(\cite{fed2009}). The simulation used fully compressive
(curl-free) forcing to excite turbulent motions, though models with
solenodial (divergence-free) forcing (not shown here) produce similar
results.  In general, compressive forcing produces greater density
contrasts with higher density enhancements and larger voids than
solenodial forcing (Federrath et al.~\cite{fed2008},
\cite{fed2009}). This could be the case for the DR21 filament since it
is here where we observe the {\sl massive dense cores} with densities
higher than 10$^5$ cm$^{-3}$ and masses up to 200 M$_\odot$.  The
velocity field primarily consists of converging flows, which create
strong density enhancements. Column density contours are overlaid on
the velocity dispersion, showing that $\sigma$ and density peaks are
offset from one another, as in our observations shown in
Fig.~\ref{infall-speed} (middle panel) of the DR21 filament. High
density gas is primarily found in regions of low $\sigma$, while lower
density regions have higher $\sigma$. We emphasise that the simulation
did not include self-gravity. Nevertheless, the velocity
dispersion--density structure is similar to what is observed in the
DR21 filament, which indicates that the observed structures can be
produced by converging supersonic gas flows only, without the
necessity of a global gravitational collapse. However, converging gas
flows are also created by global collapse. If supersonic flows and
global collapse act simultaneously, it is difficult to disentangle
their individual contribution to the creation of the velocity
dispersion--density structure seen in our observations and in the
simulations (see Sect.~\ref{largescale} below for a general discussion
of the origin of converging flows).

\subsection{Self-gravity, the main driving force in the DR21 filament?} \label{collapse} 

Observationally, the ubiquity of the blue-shifted asymmetric line
profiles in all optically thick lines indicates inflow motions. This
is a strong hint that self-gravitation plays a significant role for
the dense and massive DR21 filament. In addition, the most massive
sub-filament F3, which is a good candidate to flow towards the main
filament, seems to bend towards the center of the largest gravitation
well of the whole filament, the DR21(OH) clump (see
Sect.~\ref{subfilaments}). This bend may be indicative of a dominant
role of gravity as the flowing material approaches the main
filament. At larger distances, the effect of large-scale magnetic
fields could be the cause that keeps it in an almost east-west
direction (see Sect.~\ref{mhd}).

As shown in Sect.~\ref{flow}, we are able to explain the spatial
displacement between the maximum of the velocity dispersion and the
dense gas and the velocity field of the DR21 filament in the context
of supersonic isothermal simulations that do not include self-gravity.
However, models of isothermal, non-magnetic, but {\sl
self-gravitating} turbulence produce a velocity divergence of
typically 0.6 km s$^{-1}$ pc$^{-1}$ for a 1 pc region
(Vazquez-Semadeni et al. \cite{vaz2008}). The authors propose that the
assumption that the velocity dispersion is produced only by random
motion is incomplete and that even in the presence of driven
turbulence a part of the observed velocity dispersion is caused by
clump-scale inward motions due to gravitational collapse.  In this
large-scale inflow (LSI) scenario, the velocity field on all scales
includes a significant inflow component.

\subsection{Role of magnetic field: MHD with self-gravity modeling} \label{mhd} 

We observed several sub-filaments that are attached to the DR21
filament (Sects.~\ref{iram}).  The orientation of the
sub-filaments is not arbitrary: they follow the magnetic field lines
that run orthogonal to the NS-orientated DR21 filament (Vall\'ee \&
Fiege \cite{vallee2006}, Kirby \cite{kirby2009}).
Magnetohydrodynamic turbulence models precisely produce such
filamentary structures (Fig.~\ref{model2}) that are aligned with the
magnetic field (Fig.~\ref{model3}). Hennebelle \& Audit
(\cite{hennebelle2008b}) found that self-gravitating MHD models produce
more filamentary structures than purely hydrodynamic models.  These
numerical simulations were specifically designed for this object and the
initial conditions resemble those used in Peretto et
al. (\cite{peretto2007}). They consist of an elongated clump with an
initial aspect ratio of 2 and a density profile $\rho(r,z)=\rho_0 / (1
+ (r/r_0)^2 + (z/z_0)^2)$, where $r=\sqrt(x^2+y^2)$, $z_0= 2 r_0$, $r_0=
5$ pc and $\rho_0=500$ cm$^{-3}$.  The density at the edge of the
cloud is equal to 50 cm$^{-3}$ and is 10 times lower than the
value outside the cloud.  Turbulence is seeded initially in such
a way that the clump is approximately in virial equilibrium. The
initial temperature is 10 K and the clump is threaded by a magnetic
field parallel to the x-axis. Its intensity is proportional to the
cloud column density and the peak value is about 7 $\mu$G making the
magnetic energy about 4 times lower than the turbulent one and 5 times
higher than the thermal energy.  This simple initial configuration for
the magnetic field is suggested by the observations of Vall\'ee \& Fiege
(\cite{vallee2006}), who show that the magnetic field is indeed
perpendicular to the major axis of DR21.  The simulation was performed 
with the Ramses code (Teyssier \cite{teyssier2002}, Fromang et al.
\cite{fromang2006}).  Since the cloud occupies a small fraction (about 5
percent) of the computational domain, the following strategy is
adopted.  Initially a uniform 128$^3$ computational grid is used
(level 7). Then, all the cells of density above 10 cm$^{-3}$ and 40
cm$^{-3}$ are refined to the level 8 and 9, respectively.  This ensures 
that initially the clump is described with about $3 \times 10^6$
computing cells.  As the collapse proceeds, the resolution is adjusted
requiring at least 10 cells per Jeans length up to level 14.

In the left-hand panel of Fig.~\ref{model2}, the distribution of the
column density is shown and in the right-hand figure panel, the
velocity field vectors are overlaid on top. The column density is
smoothed to a beam of $\sim$0.1 pc angular resolution (that of the
IRAM 30m molecular line observations). The resemblance to the DR21
filament is evident with an elongated vertical main filament and
several subfilaments attached to it. Several clumps/cores (size scales
$<$0.5 pc) with high H$_2$ column densities ($>$10$^{24}$ cm$^{-2}$)
have formed in the filament, in agreement with our observations.

Figure~\ref{model3} shows the mean velocity dispersion,  the mean
velocity field along the line of sight, and the averaged magnetic
field. For the velocity dispersion $\sigma$, we
observe the same offset between regions of high density and $\sigma$
as seen in Fig.~\ref{model1}. The modeled velocity field is as
inhomogenous as the observed one, though the gradients have slightly
larger values (up to a few km s$^{-1}$ pc$^{-1}$) and the same direction (no
turn in sign as we found for the region around the DR21(OH) clump).
The magnetic field is particularly interesting. The total magnetic
intensity is displayed in Fig.~\ref{model3} (bottom), the arrows showing 
the direction of the magnetic field in the xz plane. The sub-filaments
are clearly aligned along the magnetic field and even sometimes
pinched by it. This is in full accordance with our observations and
points toward a scenario in which the DR21 filament is fed with
material along EW- orientated magnetic field lines. A full network of
these subfilaments was detected in recent JCMT $^{12}$CO 3$\to$2 data
(Richer et al., in prep.). A more detailed discussion of these
observations and MHD modelling will be presented in a
forthcoming paper (Csengeri, Hennebelle et al., in prep.).

However, the prinicipal driver of these flows remains unclear. Is it
external, large-scale driven turbulence or the gravitational
attraction of the gas itself, or both? A possible scenario is that
first, the turbulence creates the density fluctuations (clumps and
filaments) by means of locally convergent flows, and then, for some of
the more massive fluctuations,gravity takes over, making the inflow
stronger and driving the material to higher densities.  We probably
see both effects in the DR21 filament: turbulent compression and
self-gravity. The northern part is probably in an earlier stage where
the external flow is still important in shaping the velocity field
(hence the offset of the 1st and 2nd velocity moments from the density
map). In the southern part, we see a more advanced stage in which
gravity dominates and gravitational infall extends all the way to the
center (leading to the coincidence of the density and velocity
structures).

\subsection{Are the observed sub-filaments able to maintain the formation of the DR21 filament?} \label{numbers} 

We observed that several sub-filaments are attached to the DR21
filament (Sect.~\ref{iram}), possibly serving as a reservoir of gas
for further mass growth. These organized flows help to sustain high
accretion rates for times long enough to build up very massive
clumps/cores (Vazquez-Semadeni et al. \cite{vaz2009}). We note that
the DR21(OH)-clump has a mass of higher than 5000 M$_\odot$ in a
radius of $\sim$0.5 pc, which makes it one of the most massive clumps
known in the Galaxy. The F3-filament (Sect.~\ref{subfilaments}) has an
average mass input rate of 1.9$\times$10$^{-3}$ M$_\odot$ yr$^{-1}$ so
that it may contribute up to $\sim 30\,$\% of the total mass infall
rate of 5.8$\times$10$^{-3}$ M$_\odot$ yr$^{-1}$ observed for the
DR21(OH) clump as a whole ({\sl Clump 14}, see Table 3) for the next
2.1$\times$10$^6$ yr. Accounting for the other less massive
sub-filaments, which may also contribute, this indicates that the mass
flow rates are of the correct order of magnitude to explain the
formation of the DR21 filament. On the other hand, they seem
insufficient to build the present 5000 M$_\odot$ DR21(OH) clump, which
suggests that the mass flow rate must have been higher in the past and
that the process that produced the DR21 filament may be declining.

\subsection{Origin of the converging flows that 
formed the DR21 filament} \label{largescale} 

One physical scenario that intrinsically includes strong dynamics is
that of molecular clouds and complexes being formed by large-scale HI
convergent flows, which would then drive the whole dynamics of the
complex down to the star-formation scale (Klessen et
al. \cite{klessen1998}, Folini \& Walder \cite{folini2006},
Vazquez-Semadeni et al. \cite{vaz2007}, Heitsch et
al. \cite{heitsch2008}, Banerjee et al. \cite{banerjee2009}, Federrath
et al. \cite{fed2009}). Common for all 'convergent-flow' scenarios is
the basic principle that the clouds result from a pileup of material,
i.e. atomic hydrogen, from large-scale supersonic flows, caused by
energetic events such as supernovae, expanding HII-regions, galactic
spiral density waves. The driving force could be the injection of
supersonic turbulence on large scales in a pre-existing molecular
complex. Supersonic turbulence necessarily leads to organized
supersonic motions on smaller scales, and then leads to the formation
of filaments and clumps, and then stars. Finally, the self-gravity of
a pre-existing molecular complex may play a role in driving the
flows. Altogether, it is difficult to trace back the origin of the
large-scale motions observed here. But at least on the observed
scales, the motions are supersonic and are converging to form dense
structures such as the DR21 filament.  Whether a more quasi-static
evolution is possible on smaller scales or in other parts of the
complex to drive the evolution towards (high-mass) star formation
cannot be addressed by the present observations. In Csengeri et
al. (\cite{csengeri2010}), we address the kinematic state of the small
scale, densest structures in the complex, the i.e., massive dense
cores, using PdBI data (Csengeri et al. in prep).

To distinguish the formation scenarios more clarly on large scales, it
is not clear which other observational signatures can be expected. The
DR21 filament itself is already a rather dense structure and in order
to detect residues of the formation process, it is neccessary to
inspect the distribution of low-density, subsonic gas using atomic
hydrogen and low-density tracers such as $^{12}$CO. We will address
this point in a subsequent paper (Csengeri, Hennebelle et al., in
prep.).

\section{Summary and conclusions} \label{summary} 

We have presented a detailed molecular line study of the molecular ridge
containing the star-forming regions DR21 and DR21(OH). This ridge is
embedded in a large-scale network of filamentary structures, revealed
by our maps of $^{13}$CO 1$\to$0, CS 2$\to$1, and N$_2$H$^+$ 1$\to$0
emission obtained with the FCRAO. It is the most massive (around 30 000
M$_\odot$) and dense (average density $\sim$10$^4$ cm$^{-3}$) filament
within the region and is labeled by us the 'DR21 filament'.  Several
sub-filaments are linked to the DR21 filament, the most massive one
runs orthogonal to the North-South oriented ridge and has a mass
(determined from $^{13}$CO 1$\to$0) of 2600 M$_\odot$ and an average
density of 690 cm$^{-3}$. Its inferred dynamical time is
$\sim$2$\times$10$^6$ yr.

Higher angular resolution IRAM molecular line observations in HCO$^+$,
H$^{13}$CO$^+$, $^{12}$CO/$^{13}$CO 2$\to$1, C$^{34}$S, N$_2$H$^+$,
and H$_2$CO resolve the detailed structure of the DR21 filalment. The
H$^{13}$CO$^+$ 1$\to$0 data show how the sub-filament seen in
$^{13}$CO connects directly to the DR21(OH) clump. From the $^{12}$CO
2$\to$1 line mapping, we confirmed the known outflow sources DR21 and
DR21(OH) and detected three new ones, correlated with the mm-continuum
sources N53, N44, and N45 (Motte et al. \cite{motte2007}). The HCO$^+$
1$\to$0 line shows self-absorbed lines across the whole
filament. Since optically thin lines peak in the gap and the blue wing
of HCO$^+$ is more intense than the red one, we conclude that this
emission feature is due to infalling gas. The typical infall speed,
determined with a simple method described in Myers et
al. (\cite{myers1996}), is 0.6--0.8 km s$^{-1}$. A more sophisicated
non-LTE modelling of the HCO$^+$ and H$^{13}$CO$^+$ lines using the
{\sl Simline} radiative transfer code yields an infall speed of 0.5
and 0.6 km s$^{-1}$ for the northern and southern part of the DR21
filament (but excluding the DR21 region itself), respectively. 

The kinematic structure of the DR21 filament is remarkable. We
measured, using the N$_2$H$^+$, H$^{13}$CO$^+$, and HCO$^+$ maps, the 
highest values of the velocity dispersion in a vertical
column of low column density gas that is offset from the dense clumps seen in
N$_2$H$^+$. These results can be explained if the filament was
produced by turbulent flows. In this case, the densest gas has been
shocked and slowed down so that the largest velocity dispersions do
occur not in the densest regions but in the outskirts (Klessen et
al.~\cite{klessen2005}, Vazquez-Semadeni et al. ~\cite{vaz2008}). We
also observe three horizontal (west-east) velocity gradients (--0.8,
--2.2, +2.3 km s$^{-1}$ pc$^{-1}$) in the filament in which the one
with the positive value marks the location where the filament is
'falling' onto the DR21(OH) clump. This velocity pattern cannot be
explained by a single rotation of the filament along a vertical
axis.  An alternative explanation would be that at least part of the
observed motions is due to convergent flows.

By comparing our observations of the DR21 filament with a hydrodynamic
(Federrath et al. \cite{fed2009}) and a magneto-hydrodynamic
turbulence model (Hennebelle et al., Teyssier \cite{teyssier2002},
Fromang et al. \cite{fromang2006}), we infer that a very dynamic and
fast mode of star formation occurs in the filament, in which gas is
continuously replenished by subfilaments attached to the main
filament. These subfilaments are aligned with the magnetic field
direction that is perpendicular to the DR21 filament (Vallee \&
Fiege \cite{vallee2006}). The DR21 filament is globally
collapsing. All our observational findings are incompatible with the
view of a quasi-static, pressure-bounded clump scenario.

\begin{acknowledgements}
We thank A. Belloche and D. Poelman for useful discussions on line
radiative transfer modelling. \\
A part of this work was supported by the
French Agence National de la Recherche (ANR) project PROBeS
No. 08-blanc-0241. T. Csengeri acknowledges support from the FP6
Marie-Curie Research Training Network 'Constellation: the origin of
stellar masses' (MRTN-CT-2006-035890). \\
R.S.K.\ acknowledges financial support from the German {\em  
Bundesministerium f\"{u}r Bildung und Forschung} via the ASTRONET  
project STAR FORMAT (grant 05A09VHA) and from the {\em Deutsche  
Forschungsgemeinschaft} (DFG) under grants no.\ KL 1358/1, KL 1358/4,  
KL 1359/5, KL 1358/10, and KL 1358/11. R.S.K.\ furthermore thanks for  
subsidies from a Frontier grant of Heidelberg University sponsored by  
the German Excellence Initiative and for support from the {\em  
Landesstiftung Baden-W{\"u}rttemberg} via their program International  
Collaboration II (grant P-LS-SPII/18). R.S.K. also thanks the KIPAC at  
Stanford University and the Department of Astronomy and Astrophysics  
at the University of California at Santa Cruz for their warm  
hospitality during a sabbatical stay in spring 2010.
\end{acknowledgements}

\appendix 

\section {Calculation of physical parameters} \label{formulas} 

\subsection {Abundances} \label{abundance} 

We use a standard [$^{12}$CO]/[$^{13}$CO] ratio of 67 (Lucas et al. \cite{lucas1998}). 

The [HCO$^+$]/[H$_2$] abundance in high-mass star-forming regions is
typically quoted to be 1.5--2.5 $\times$ 10$^{-9}$ (Peretto et
al.~\cite{peretto2006}, Marseille et al.~\cite{marseille2008}). For
our mass determination from H$^{13}$CO$^+$, we apply the abundance
value obtained from the {\sl Simline}  modelling (Sect.~\ref{simline}), which
is 2.5 $\times$ 10$^{-9}$. 
 
\subsection {Determination of mass from molecular lines} \label{mass-det}   

The total column density of any optically thin molecule can be determined from 
\begin{eqnarray}
N [cm^{-2}] & = & f(T_{ex}) \int T_{mb} [K] d{\rm v} [km s^{-1}]
\nonumber \\ 
\end{eqnarray}
where 
\begin{eqnarray}
f(T_{ex}) & = & \frac{3hZ}{8 \pi^3 \mu^2 J_t}
\frac{\exp(h\nu/kT_{ex})}{[1-\exp(-h\nu/kT_{ex})] (J(T_{ex}) - J(T_{BG}))} 
\nonumber \\  
& &   
\label{ncol}  
\end{eqnarray}
and the Partition function is given by 
\begin{eqnarray}
Z & = & \frac{2kT_{ex}}{h \nu}+1/3 
\end{eqnarray}
in which $h$ and $k$ denote the Planck and the Boltzman constants,
respectively, $\nu$ is the frequency [GHz], $\mu$ is the dipole moment [Debye], 
$J_t$ is the upper value of the rotational quantum number and $\int T_{mb} dv$
is the velocity integrated line intensity on a main beam temperature scale 
\begin{eqnarray}
J(T_{ex}) & = & \frac{h\nu}{k(\exp(h\nu/(kT_{ex})-1)} 
\end{eqnarray}
and $J(T_{BG}) = J(2.7K)$. \\

\noindent The values for the {\bf H$^{13}$CO$^+$ 1$\to$0} transition are $\nu$=86.75433 GHz,
h$\nu$/k=4.163 K, $\mu$=3.3 Debye, $J_t$=1, and $J(2.7K)$=1.13. The
temperature-dependent factor f(T$_{ex}$) is 
(2.43, 2.84, 4.39, 9.4)$\times$10$^{12}$ for 10, 20, 40, and 100 K, respectively.

The values for the {\bf $^{13}$CO 1$\to$0} transition are $\nu$=110.201 GHz,
h$\nu$/k=5.29 K, $\mu$=0.112 Debye, $J_t$=1, and $J(2.7K)$=0.868. The
temperature-dependent factor f(T$_{ex}$) is 
(0.91,1.25,1.66,2.08)$\times$10$^{15}$ for 10, 20, 30, and 40 K, respectively. \\


The H$_2$ column density from H$^{13}$CO$^+$ is calculated to be 
\begin{eqnarray}
{\rm N(H}_2) [{\rm cm}^{-2}] & = & 2.43 \, \times \, 10^{10}\, {\rm N(H}^{13}{\rm CO}^+)  
\,\,\,\,\, \nonumber \\ 
\end{eqnarray}
and for $^{13}$CO 1$\to$0 from 
\begin{eqnarray}
{\rm N(H}_2) [{\rm cm}^{-2}] & = & 4.7 \,\times \, 10^{5}\, {\rm N(}^{13}{\rm CO}) \,\,\,\,\,\nonumber \\ 
\end{eqnarray}
The final mass is determined to be  
\begin{eqnarray}
{\rm M} [M_\odot] & = & {\rm N(H}_2) \, d^2  \,A (\pi /180)^2 \, 2m_H \, \mu , 
\nonumber \\ 
\end{eqnarray}
where the projected area $A$ is in square degrees, the distance $d$ in parsec, the mass of 
hydrogen m$_H$, and $\mu$=1.36.


\begin{thebibliography}{} 

\bibitem[1992]{adler1992} 
 Adler, D.S., Lo, K.Y., Wright, M.C.H. et al., 1992, ApJ, 392, 497 


\bibitem[2009]{audit2009} 
 Audit, E., Hennebelle, P., 2009, A\&A in press, arXiv:0911.0748

\bibitem[1999]{ball1999} 
 Ballesteros-Paredes, J., Vazquez-Semadeni, E., Scalo, J., 1999, ApJ, 512, 286 

\bibitem[2003]{ball2003} 
 Ballesteros-Paredes, J., Klessen, R., Vazquez-Semadeni, E., Scalo, J., 2003, ApJ, 592, 188

 

\bibitem[2007]{banerjee2007} 
 Banerjee, R., Pudritz, R.E., 2007, ApJ, 660, 470 

\bibitem[2009]{banerjee2009} 
 Banerjee, R., Vazquez-Semadeni, E., Hennebelle P., Klessen, R.S., 2009, MNRAS, 398, 1082  

\bibitem[2009]{bate2009} 
 Bate, M., 2009, MNRAS, 392, 1363 

\bibitem[1988]{batrla1988} 
 Batrla, W., Menten, K.M., 1988, ApJ, 329, L117

\bibitem[1982]{bieging1982} 
 Bieging, J.H., Wilson, T.L., Downes, D., 1982, A\&AS, 49, 607


\bibitem[2010]{bontemps2010} 
 Bontemps, S., Motte, F., Csengeri, T., Schneider, N., 2010, A\&A submitted,
 arXiv:0909.2315

\bibitem[2006]{bonnell2006} 
 Bonnell I., \& Bate M., 2006, MNRAS 370, 488 

\bibitem[2003]{brunt2003} 
 Brunt, C.M., 2003, ApJ, 583, 280

 
\bibitem[1993]{chandler1993} 
 Chandler, C.J., Gear, W.K., Chini, R., 1993, MNRAS, 260, 337

\bibitem[2010]{csengeri2010}
 Csengeri, T., Bontemps, S., Schneider, N., et al., submitted to A\&A



 

\bibitem[2007]{davis2007} 
 Davis, C.J., Kumar, M.S.N., Sandell, G., et al., 2007, MNRAS, 374, 29

 
\bibitem[1978]{dickel1978} 
 Dickel, H.R., \& Wendker, H.J., 1978, A\&A, 66, 289 

\bibitem[1978]{dickeldickel1978} 
 Dickel, J.R., Dickel, H.R., \& Wilson, W.J., 1978, ApJ, 223, 840 

\bibitem[1966]{downes1966} 
 Downes, D., \& Rinehart, R., 1966, ApJ, 144, 937 


\bibitem[2008]{fed2008} 
 Federrath, C., Klessen, R., Schmidt, W., 2008, ApJ, 688, L79 

\bibitem[2009]{fed2009} 
 Federrath, C., Roman-Duval, J., Klessen, R., Schmidt, W., \& Mac Low, M.-M., 
 2009, A\&A accepted, arXiv:0905.1060 

\bibitem[2006]{folini2006} 
 Folini, D., Walder, R., 2006, A\&A, 459, 1 

\bibitem[2006]{fromang2006} 
 Fromang, S., Hennebelle, P., Teyssier, R., 2006, A\&A, 457, 371



\bibitem[1991a]{garden1991a} 
 Garden, R.P., Geballe, T.R., Gatley, I., et al., 1991, ApJ, 366, 474

\bibitem[1991b]{garden1991b} 
 Garden, R.P., Hayashi, M., Gatley, I., et al., 1991, ApJ, 374, 540

\bibitem[1992]{garden1992} 
 Garden, R.P., \& Carlstrom, J.E., 1992, ApJ, 392, 602

\bibitem[1977]{genzel1977}
 Genzel, R., Downes, D., 1977, A\&AS, 30, 145


\bibitem[1974]{goldreich1974}
 Goldreich, P, Kwan, J., 1974, ApJ, 189, 441

\bibitem[2007]{gomez2007}
 Gomez, G.C., Vazquez-Semadeni, E., Shadmehri, M., Ballesteros-Paredes, J., 2007, ApJ, 669, 1042

\bibitem[1993]{goodman1993}
 Goodman, A.A., Benson, P.J., Fuller, G.A., Myers, P.C., 1993, ApJ, 406, 528


\bibitem[1973]{harris1973}
 Harris, S., 1973, MNRAS, 162, 5P

\bibitem[2008]{heitschhart2008}
 Heitsch, F., Hartmann, L.W., 2008, ApJ 689, 290

\bibitem[2008]{heitsch2008}
 Heitsch, F., Hartmann, L.W., Slyz, A.D. et al., 2008, ApJ 674, 316

\bibitem[2007]{hennebelle2007}
 Hennebelle, P., Audit, E., 2007, A\&A, 465, 431 

\bibitem[2008]{hennebelle2008a}
 Hennebelle, P., Teyssier, R., 2008, A\&A, 477, 25 


\bibitem[2008]{hennebelle2008b}
 Hennebelle, P., Banerjee, R., Vazquez-Semadeni, E., Klessen, R., Audit, E., 2008, 486, L43

\bibitem[2008]{hennebelle2008c}
 Hennebelle, P., Chabrier, G., 2008, ApJ, 684, 395 

\bibitem[2009]{hora2009}
 Hora, J., Bontemps, Megeath, T., Schneider, N., et al., 2009, AAS 213, 356.01, 
 vol. 41, p.498

\bibitem[2007]{jakob2007} 
 Jakob, H., Kramer, C., Simon, R., Schneider, N., Ossenkopf,
 V., Bontemps, S., et al., 2007, A\&A, 461, 999


\bibitem[2009]{kirby2009}
 Kirby, L., 2009, ApJ, 694, 1056

\bibitem[1998]{klessen1998}
 Klessen, R.S., Burkert, A., Bate, M.R.,1998, ApJ, 501, 205

\bibitem[2000]{klessen2000}
 Klessen, R.S., Heitsch, F., Mac Low, M.-M., 2000, ApJ, 535, 887

\bibitem[2001]{klessen2001}
 Klessen, R.S., 2001, ApJ, 550, L77

\bibitem[2005]{klessen2005}
 Klessen, R.S., Ballesteros-Paredes, J., Vazquez-Semadeni, E., et al., 2005, ApJ, 620, 786

\bibitem[2010]{klessen2010}
 Klessen, R.S., Hennebelle, P., 2010, submitted to A\&A, arXiv:0912.0288

\bibitem[2000]{knoedl2000}
 Kn\"odlseder, J., 2000, A\&A, 360, 539


 
\bibitem[2007]{kumar2007} 
 Kumar, M.S.N., Davis, C.J., Grave, J.M.C., et al., 2007, MNRAS, 374,
 54 
 

\bibitem[2005]{krumholz2005}
 Krumholz, M., McKee, C.F., Klein, R.I., 2005, ApJ, 618, L33

\bibitem[2006]{krumholz2006}
 Krumholz, M., 2006, ApJ, 641, L45

 

\bibitem[2003]{lee2003} 
 Lee, J.E., Evans, N.J., Shirley, Y.L., 2003, ApJ, 583, 789

\bibitem[1998]{lucas1998}
 Lucas, R., Liszt, H., 1998, A\&A, 337, 246 


\bibitem[2004]{maclow2004} 
 Mac Low, M.-M., Klessen, R., 2004, Reviews of Modern Physics, vol.76, Issue 1, 125-194 

\bibitem[1991]{mangum1991}
 Mangum, J.G., Wooten, A., Mundy, L.G., 1991, ApJ, 378, 576 

\bibitem[1992]{mangum1992}
 Mangum, J.G., Wooten, A., Mundy, L.G., 1992, ApJ, 388, 467 

\bibitem[2008]{marseille2008}
 Marseille, M., Bontemps, S., van der Tak, F., Herpin, F., Purcell,C.R., 2008, A\&A, 488, 579

\bibitem[2004]{marston2004} 
 Marston, A.P., Reach, W.T., Noriega-Crispo, A., et al., 2004, ApJS,
 154, 333

\bibitem[1986]{mauers1986} 
 Mauersberger, R., Henkel, C., Wilson, T.L., Walmsley, C.M., 1986,
 A\&A, 162, 199

\bibitem[2003]{mckee2003}
 McKee, C., Tan, J., 2003, ApJ, 585, 850 

\bibitem[2007]{mckeeostriker2007} 
 McKee, C.F., Ostriker, E.C., 2007, Annu.Rev.Astron.Astrophys. 45, 565

  
\bibitem[2003]{motte2003} 
 Motte, F., Schilke P., Lis, D., 2003, ApJ, 582, 277

\bibitem[2007]{motte2007} 
 Motte, F., Bontemps, S., Schilke P., Schneider, N., Menten, K., 2007,
 A\&A, 476, 1243

\bibitem[1996]{myers1996}
 Myers, P.C., Mardones, M., Tafalla, M., et al., 1996, ApJ, 465,L133

\bibitem[2009]{myers2009}
 Myers, P.C., 2009, ApJ, 700, 1609

\bibitem[1982]{norris1982} 
 Norris, R.P., Booth, R.S., Diamond, P.J., Porter, N.D., 1982, MNRAS, 201, 191

\bibitem[1983]{nyman1983} 
 Nyman, L.A., 1983, 120, 307 


\bibitem[2001]{ossenkopf2001}
 Ossenkopf, V., Trojan, C., Stutzki, J., 2001, A\&A, 378, 608


\bibitem[2002]{padoan2002} 
 Padoan, P., Nordlund, A. 2002, ApJ, 576, 870

\bibitem[2006]{peretto2006} 
 Peretto, N., Andr\'e, P., Belloche, A., 2006, A\&A, 445, 979 

\bibitem[2007]{peretto2007} 
 Peretto, N., Hennebelle, P., Andr\'e, P., 2007, A\&A, 464, 983

\bibitem[2010]{peters2010} 
 Peters, T., Banerjee, R., Klessen, R., et al., 2010, ApJ, 711, 1017


\bibitem[1990]{plambeck1990} 
 Plambeck, R.L., \& Menten, K., 1990, ApJ, 364, 555
 
\bibitem[2008]{reipurth2008} 
 Reipurth, B., Schneider, N., 2008, Handbook of star forming regions, ASP 
 

\bibitem[1994]{richardson1994} 
 Richardson, K.J., Sandell, G., Cunningham, C.T., et al., 1994, A\&A, 286, 555

\bibitem[1992]{russell1992} 
 Russell, A.P.G., Bally, J., Padman,R., Hills, R.E., 1992, ApJ, 387, 219
 
 
 
 
 
\bibitem[2006]{schneider2006} 
 Schneider, N., Bontemps, S., Simon, R., Jakob, H., Motte, F.,  Miller, M., 
 Kramer, C., Stutzki, J., 2006, A\&A, 458, 855 

\bibitem[2007]{schneider2007} 
 Schneider, N., Simon, R., Bontemps, S., Comer\'on, F., Motte, F.,
 2007, A\&A, 474, 873

\bibitem[2010]{schneider2010} 
 Schneider, N., Bontemps, S., Simon, R., et al., 2010, submitted to A\&A, arXiv:1001.2453

\bibitem[2004]{shepherd2004} 
 Shepherd, D.S.,  Kurtz, S.E., Testi, L., 2004, ApJ, 601, 952


 

 

\bibitem[2002]{tafalla2002} 
 Tafalla, M., Myers, P.C., Caselli, P., Walmsley, C.M., Comito, C.,
 2002, ApJ, 569, 815

\bibitem[2002]{teyssier2002}
 Teyssier, R., 2002, A\&A, 385, 337

 
\bibitem[2006]{vallee2006} 
 Vall\'ee, J.P. \& Fiege, J.D., 2006, ApJ, 636, 332  

\bibitem[2002]{vaz2002} 
 Vazquez-Semadeni, E., Shadmehri, M., Ballesteros-Paredes, J., 2002, sub. to ApJ, 
 arXiv:0208245

\bibitem[2007]{vaz2007} 
 Vazquez-Semadeni, E., Gomez, G.C., Jappsen, A.K.,et al., 2007, ApJ, 657, 870

\bibitem[2008]{vaz2008} 
 Vazquez-Semadeni, E., Gonzales, R.F., Ballesteros-Paredes, J., et al., 2008, MNRAS, 390, 769

\bibitem[2009]{vaz2009} 
 Vazquez-Semadeni, E.,  Gomez, G.C., Jappsen, A.K.,et al., 2009, ApJ, 707, 1023


\bibitem[1990]{wilson1990} 
 Wilson, T.L. \& Mauersberger, R. 1990, A\&A, 239, 305

\bibitem[2008]{wienen2008} 
 Wienen, M., 2008, Diplomarbeit, University of Bonn

\bibitem[2002]{yorke2002} 
 Yorke, H.W., Sonnhalter, C., 2002, ApJ, 569, 846

\end{thebibliography}
\end{document}